# Action Research Can Swing the Balance in Experimental Software Engineering

Paulo Sérgio Medeiros dos Santos            Guilherme Horta Travassos

System Engineering and Computer Science Department
COPPE/Federal University of Rio de Janeiro
P. O. Box 68511 – 21941-972 Rio de Janeiro – RJ – Brazil
{pasemes, ght}@cos.ufrj.br

## Table of Contents

**Abstract**

*In general, professionals still ignore scientific evidence in place of expert opinions in most of their decision-making. For this reason, it is still common to see the adoption of new software technologies in the field without any scientific basis or well-grounded criteria, but on the opinions of experts. Experimental Software Engineering is of paramount importance to provide the foundations to understand the limits and applicability of software technologies. The need to better observe and understand the practice of Software Engineering leads us to look for alternative experimental approaches to support our studies. Different research strategies can be used to explore different Software Engineering practices. Action Research can be seen as one alternative to intensify the conducting of important experimental studies with results of great value while investigating the Software Engineering practices in depth. In this chapter, a discussion on the use of Action Research in Software Engineering is presented. As indicated by a technical literature survey, along the years a growing tendency for addressing different research topics in Software Engineering through Action Research studies has been seen. This behaviour can indicate the great potential of its applicability in our scientific field. Despite their clear benefits and diversity of application, the initial findings also revealed that the rigour and control of such studies should improve in Software Engineering. Aiming at better explaining the application of Action Research, an experimental study (**in vivo**) on the investigation of the subjective decisions of software developers, concerned with the refactoring of source code to improve source code quality in a distributed software development context is depicted. A software engineering theory regarding refactoring and some guidance on how to accomplish an Action Research study in Software Engineering supplement the discussions in this chapter.*

**Keywords**: *Action Research, Refactoring, **in vivo** Study, Software Engineering Theory, Scientific Knowledge Management, Experimental Software Engineering*

## 1. Introduction

Software Engineering is a multidisciplinary discipline involving different social and technological features. To understand how software engineers maintain complex software systems, it is necessary to investigate not only the tools and processes they use, but also the cognitive and social processes surrounding them. It requires the study of human activities. We need to understand how software engineers individually develop software as well as how teams and organizations coordinate their efforts (Easterbrook *et al.*, 2008).

One possible approach to support this understanding is experimentation. Experimental Software Engineering is of paramount importance to provide the foundations to understand the limits and applicability of software technologies. It intends to make the practice of Software Engineering predictable and economically feasible. Indeed, the importance of experimentation has grown significantly over the recent years as many technical literature surveys and researchers have pointed out, such as Sjöberg *et al.* (2005), Basili and Elbaum (2006), Höfer and Tichy (2007), and Kampenes *et al.* (2009).

However, despite all the perceived importance, experimentation still does not seem to have achieved broad use and expansion in Software Engineering as other science fields demonstrate today, e.g. Medicine (Charters et al., 2009). For instance, some years ago Tichy (1998) had pointed out that there was no solid evidence in comparing basic technologies for the practice of Software Engineering such as the paradigms of functional and object-oriented development, present in almost all industrial software projects. Both certainly have advantages and disadvantages. And, in an attempt to show such technologies to the Software Engineering community, demonstrations and heuristics were usually developed to support their selection and use. However, demonstrations rarely produce solid evidence. Despite all the effort invested to reveal evidence about different software technologies other than development paradigms, the use of these kinds of evaluation can still be seen in the field. Different factors can be attributed to this limited application of experimentation in Software Engineering. For example, the complexity, cost, and risks concerned with different types of experiments apparently represented obstacles to its adoption (Juristo and Moreno, 2001).

The understanding of software engineering researchers on the importance of experimentation had promoted the first and relevant experiences with the experimental method in Software Engineering. According to Basili *et al.* (1986), those experiments were strongly influenced by areas such as Physics and Medicine which, due to their level of maturity, focused on controlled studies. This type of study can be difficult to control in real software projects as the large number of unknown context variables may present a high degree of risk to these experiments (Basili, 1996). Additionally, experimental studies can be time-consuming and generate large volume of information making the management of scientific knowledge difficult (Shull *et al.*, 2001).

Later, the lessons learned with controlled studies were described in books dedicated exclusively to the topic (Wöhlin *et al.*, 2000, Juristo and Moreno, 2001). Based on this experience, researchers started to investigate how to enlarge the experimentation opportunities by introducing new methods and research strategies such as case studies, ethnography, surveys and interview techniques (Harrison *et al.* 1999, Wöhlin *et al.* 2003, Zelkowitz, 2007, Easterbrook *et al.* 2008) in the field.

A motivation for the search for new research approaches relates to the fact that although controlled studies and measurement allow the observation of relationships among variables by means of statistical tests, their limitations may restrict what one can see and, therefore, investigate (Pfleeger, 1999). Similarly, years later, Kitchenham (2007) stated that the excessive emphasis on using controlled studies in Software Engineering 'may place ourselves examining phenomena that are a result of abstracting the technology away from its usage context and not the characteristics of the technology itself.' In other words, the premature insistence on accuracy can inhibit the progress and lead scientists to formulate problems in ways that can be measured but have limited relevance in relation to the characteristics of the problem (Argyris *et al.*, 1985). For this reason, researchers run the risk of considering the professionals from industry responsible for misusing their proposed software technologies when the actual problem is theirs, in failing to understand the complexity of the context under which the technologies will be used (Kitchenham, 2007).

The need to better observe and understand the practice of Software Engineering leads us to look for alternative experimental approaches. Consequently, this takes us to one of the core Software Engineering research concerns: the relevance of scientific results. Relevance relates to the usefulness of the results in an industrial context (e.g., providing results that enable the development of guidelines) as well as in the academic context (e.g., allowing a better understanding of a phenomenon through the construction of theories) (Sjöberg *et al.*, 2007). Other researchers have also emphasized this need, such as Kitchenham *et al.* (2004) and Glass (2009). The industry itself shows the importance of increasing the relevance of studies in Software Engineering. Recent works (Rainer et al., 2005, Kitchenham et al., 2007) suggest that professionals ignore scientific evidence in place of expert opinion in most of their decision-making. For this reason, it is still common that new software technologies are adopted without any scientific basis or well-grounded criteria, and the opinions of experts may be limited as their experience is contained in a limited set of options.

Therefore, different research strategies should be used to explore the different features in Software Engineering practice. The methodology in Action Research can be seen as an alternative to intensify the execution of relevant studies and the acquisition of great value results (Baskerville and Wood-Harper, 1996) while allowing for in-depth investigation of the practices in Software Engineering.

Action Research has its origins associated with the early interventionist practices carried out by Kurt Lewin in the 1940s in the course of social-technical experiments. The initial stimulus for the rise and design of the main action research objectives and aspirations came from a generalized difficulty at that time in translating the results of social research into practical actions (Carr, 2006).

The history of Action Research is usually split into at least two stages (Reason and Bradbury, 2001). The first stage relates to Kurt Lewin's initial practices, and the second one is associated with the resurgence of interest in educational research on Action Research in the early 1970s, after its initial rejection due to the predominance of a positivist stance in the Social Sciences. Some reasons for this renewed interest in Action Research include, for example, the claim that the professionalization of teachers should be improved by also giving them a researcher role to allow assessing curricular guidelines in the classroom and improving teaching practices (Carr, 2006). Thus, seen in this way, Action Research was transformed from a method through which professionals applied the scientific theories of Social Sciences in practice to a method that allowed practitioners to assess the adequacy of their own tacit theories in their practice. Nowadays, Action Research is also used in other scientific areas such as management, nursing, and information systems (Dick, 2004) to support the role of researching in their daily activities.

This brief description of the history of Action Research shows how the methodology has evolved through the many changes caused by new interpretations and uses researchers from different science fields gave it (Burns, 2005). A clear example of this continuous transformation process is represented by the work of Baskerville and Wood-Harper (1998). Researchers were able to identify a large diversity of Action Research processes, represented by 10 different 'formats' used in the area of Information Systems. Fortunately, apart from all of these proposals, a comprehensive Action Research process (based on the 1978 Susman and Evered model) can be identified, that can support the accomplishment of Action Research studies in Software Engineering (Davison *et al.*, 2004).

As it will be described in this chapter, Action Research is starting to support studies in Software Engineering. Its characteristics suggest its application can benefit research in the field, as it simultaneously allows the performing of research and action. The action is usually associated with some transformation in a community, organization or project, while the research is characterized by a wide understanding of a transformation phenomenon by the researcher (research community), person (client), or both. Avison *et al*. (1999) emphasizes that Action Research regards 'more what practitioners do than what they say they do'. According to this claim, Sjöberg *et al*. (2007) have pointed the Action Research methodology as 'the kind of study where the most realistic scenario is found' as it involves a real industry context to investigate the results of concrete actions.

The social challenges dealt with by researchers in Software Engineering investigations make Action Research an useful research methodology due to its characteristics and possibility of obtaining relevant results. We found out that there is an increasing tendency on using Action Research in Software Engineering to address different research topics indicating a great potential for its applicability in the area. The initial findings also revealed that the rigour and control of Action Research studies should be improved in Software Engineering related studies. This can stem from borrowing a methodology from other scientific domain (Social Sciences) where studies are usually described using a different way of communicating and exchanging thoughts. Thus, even if there are few studies using the methodology of Action Research in Software Engineering a discussion of the particularities of the methodology in the area seems necessary, furthermore considering the increasing interest in its use.

Aiming at addressing these issues, we conducted two Action Research studies. Besides the expected benefits to the software project, the aim is to observe how action research can be applied in Software Engineering. The first study investigated a checklist-based inspection technique to improve the comprehensibility of use case models in a real software project. This study was first planned to strictly follow Action Research procedures. However, our previous experiment on accomplishing controlled studies concerned with software inspections influenced us to rely excessively on quantitative data, although the study was indeed an Action Research based study. Therefore, we decided to better understand Action Research to improve our investigation capabilities. So, in Section 2 the reader can find descriptions of the main concepts of Action Research, as well as its limits and advantages, comparing it to other methodologies and showing its process and core principles. After that, to supplement this overview, Section 3 aims at illustrating the results of a technical literature survey conducted to evaluate the current degree of use of Action Research in Software Engineering.

Having become better acquainted with the methodology, the second study spontaneously emerged from a problematic situation related to source code quality in a distributed software development context which the authors participated in at that time. The second study examined the subjective decision concerned with source code refactoring. Due to its richer scenario and adherence to the Action Research methodology, including the use of qualitative data techniques (i.e., grounded theory), this will be tackled in Section 4. However, the working experiences in both studies will underlie our suggestions, recommendations, and practices on how to apply and when to choose Action Research as a research methodology in Software Engineering (in Section 5). Finally, in Section 6, some conclusions are given, including additional remarks intending to strengthen this intended chapter argument: Action Research can swing the balance to support the revealing of evidence in Experimental Software Engineering.

## 2. Action Research Overview

Based on the understanding of how Action Research is positioned within scientific research paradigms this section details many aspects of the methodology and describes a canonical process identified in the technical literature. To understand the objectives, limitations and benefits of research strategies, the main criteria distinguishing them are going to be addressed in this section. For the sake of simplicity, there is no intention to go into excessive detail in any of them.

### 2.1. Background

The distinction between research strategies can be made by exploring the paradigm concept. A paradigm can be seen as a set of basic truths (whose veracity we cannot ascertain) dealing with primary principles (axioms and doctrines). This set represents a worldview defining the nature of the 'world', the place of each individual, and a set of possible relationships with the world and its parts (Guba and Lincoln, 1994). For Hathaway (1995), scientific paradigms can essentially act as a lens through which researchers can observe and understand the problems in their areas and produce scientific contributions to them. The scientific paradigms impose what researchers consider as data, what their role in scientific research is, what to consider knowledge and how reality is seen and accessed. In short, scientists bring about their daily assumptions based on the knowledge, reality, and research strategies they hold. Scientific paradigms are usually marked by ontology, epistemology concepts, and methodology (Guba and Lincoln, 1994). Epistemology relates to the nature of human knowledge and how it is understood and communicated, ontology deals with the basic issues of the nature of the world, regardless of who tries to observe it, and methodology addresses how (which methods and approaches) individuals acquire knowledge on the world .

The four predominant scientific paradigms found in the technical literature (Guba and Lincoln, 1994, Healy and Perry, 2000, Easterbrook *et al.*, 2008) can be summarized as follows:

- **Positivism** states that all knowledge should be based on logical inference from a set of observable facts. Positivists are reductionists because they study the events by breaking them into simpler components. This corresponds to their belief that scientific knowledge can be built incrementally from verifiable observations and the inferences based on them. In other words, data and analysis are neutral and do not change due to the fact that they are being observed. However, a positivist view is usually inappropriate to address social phenomena when they involve humans and their experiences, as it ignores the ability to reflect on the problems and to act on them in an inter-dependent approach. The positivist paradigm is closely related to the controlled study strategy; however, surveys and case studies can also be conducted under this paradigm.

- **Constructivism**, also known as interpretivism, rejects the idea of separating scientific knowledge from its human context. Constructions are not more or less absolutely 'true' but simply more or less well-formed and/or sophisticated, better informing on the complexity of the object studied. Constructivists focus less on theory verification and more on understanding how different people grasp the world and how they attribute meaning to their actions. Theories may arise in this process but always linked to the studied context producing the so-called local theory. Constructivism is typically associated with ethnography but can often be related to exploratory case studies, Action Research (Ludema *et al.*, 2001 *apud* Reason and Bradbury, 2001), and surveys.

- **Critical theory** considers scientific knowledge from its ability to release people from schools of thought. Critical theorists argue that research is a political act as knowledge gives power to different groups in society or strengthens existing authority power structures. In this sense, they choose what to research based on who will help and prefer the participatory approaches or supporting roles. Therefore, it is assumed that the investigator and investigated object are interactively linked, with the values of the researcher inevitably influencing the investigation. In Software Engineering this paradigm can be associated, for instance, to the software process improvement and agile methods communities. Moreover, critical theorists often make use of case studies to draw attention to something that needs to be modified. However, it is Action Research that best reflects this paradigm (Carr and Kemis, 1986 *apud* Reason and Bradbury, 2001).

- **Pragmatism** recognizes that all knowledge is incomplete and estimated at certain level, and its value depends on the acquiring method. For pragmatists, knowledge is judged by how useful it can be to solve practical problems. Therefore, there is a degree of relativism: what is useful for one person may not be for another one, and thus truth is relative to the observer. To overcome this criticism, pragmatists emphasize the importance of consensus as the external 'guardian' of objectivity; this can be seen for instance in the case of tradition/practice (are the results consistent with current knowledge?) and community (publishers, judges, and professionals). In short, the pragmatist adopts the engineering approach to research and emphasize practical rather than abstract knowledge, using any appropriate method to obtain it. One method often used, especially when there is no *a priori* known solution, is Action Research, as it allows the combination of different research strategies and data collection techniques (Greenwood and Levin, 1998 *apud* Reason and Bradbury, 2001).

These scientific paradigms can be classified into two main approaches to obtain and construct knowledge: quantitative and qualitative. Positivism is related to the quantitative approach whilst the other paradigms to the qualitative one (Healy and Perry, 2000). A scientific paradigm can use both approaches, but always focuses more on one of them depending on its main orientation.

The quantitative approach aims at measuring and analyzing the causal relationships between variables representing the observed object characteristics. It operates with data in numerical form, collected from a representative sample, and usually analyzed using statistical methods. This way, the main goal is to identify the independent and dependent variables, reducing problem complexity so that the initial hypotheses can be confirmed or refuted (Wöhlin *et al.*, 2000). The main disadvantages of this approach concern the fact that its objectivist and reductionist nature limits a detailed (descriptive) understanding of the real world properties and characteristics, so that the observation of part of reality cannot be accomplished without losing the phenomenon importance in the whole (Krauss, 2005).

The qualitative approach, on the other hand, is best described by evaluating how it differs from the quantitative one (Näslund, 2002): (1) it approximates the investigated perspective through interviews and observation, (2) it tends to emphasize detailed descriptions while the quantitative approach is less concerned with this kind of detail, (3) qualitative data is represented not only by numbers but by text and images too. For Seaman (1999), the main advantage of using a qualitative approach is that it requires the researcher to dive into the complexity of the problem instead of abstracting it. Thus, the qualitative approach is often used to answer 'how/why' a causality phenomenon revealed by quantitative approaches occurred. However, the drawback that generated knowledge can be considered more 'vague' and 'fluid', especially in a technical community such as that of Software Engineering (Seaman, 1999). Furthermore, this makes results more difficult to aggregate or simplify.

To understand how the qualitative and quantitative approaches relate to the aforementioned scientific paradigms, the graph in Figure 1 illustrates the intensity with which the four scientific paradigms make use of these approaches. It is interesting to observe, considering all the paradigms where Action Research appears as an option, its wide use spectrum, ranging from an observer role by the researcher in constructivism, to a facilitator position in critical theory and reaching a more active problem solving stance in pragmatism (Burns, 2005). For this reason, it is not always possible to classify each action research study into one paradigm, even if the researcher adopts one position as a foundation. But what is best defined in Action Research as a paradigm is a non-positivist stance, i.e. a non artificial separation between the observer and that which is observed.

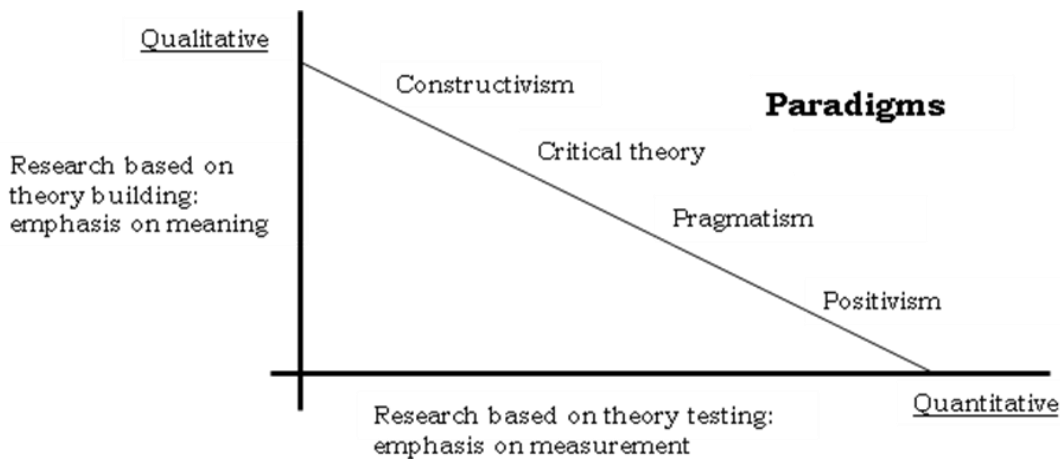

**Figure 1 - Scientific paradigms and the qualitative and quantitative approaches – evolved from Healy and Perry (2000)**

## 2.2. Action Research Process

Action Research can be defined as 'a kind of social research with experimental basis that is conceived and conducted in close association with an action or a collective problem resolution where researchers and participants are involved in a cooperative way' (Thiollent, 2007). In terms of process, the essence behind this definition can be thought into two stages (Figure 2). The first stage involves researchers and participants who collaboratively analyze a problem situation. Based on this examination, they formulate theories meant to explain the circumstances faced. These theories are in fact conjectures and speculative practical knowledge, which are translated to a research topic later on. The second stage involves what we can call collaborative practical experiments (attempts). In this stage, changes are introduced and their effects studied. These two stages are iterated until the problem can be solved.

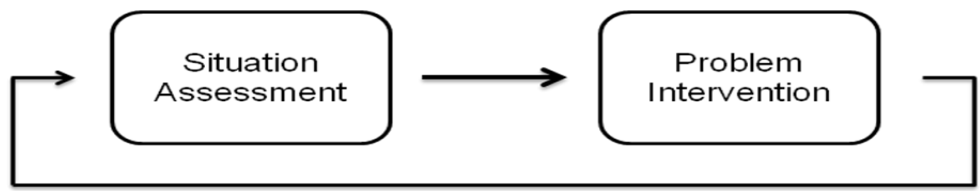

**Figure 2 - Action research two-stage simplified process – adapted from Baskerville (2007)**

Argyris *et al.* (1985) makes an analogy between experimentation and a process similar to the one shown in Figure 2 saying that the practical knowledge embedded in the action (represented by the identification of causal factors that can be manipulated to get the desired consequences within a set of circumstances) is the hypothesis being tested. If the intended consequences occur, then the hypothesis is confirmed. Otherwise it is rejected (or the alternative hypotheses based on the supposed environmental conditions can be accepted).

Even when there is no neutrality on the observation and manipulation of the study environment, which is required for hypothesis refutation, any scientific methodology should contribute to the body of scientific knowledge in a consistent and rigorous way. Moreover, the scientific evaluation of any proposition within the body of scientific knowledge has to be replicable. As a result, an investigation process similar to that shown in Figure 2 is rarely adopted directly in practice as it does not define any kind of guidance regarding these issues.

According to Checkland and Holwell (1998), any research methodology can be linked to the following elements: a cohesive set of ideas forming a conceptual framework *F* that is used in a methodology *M* to investigate an area of concern *A*. In Software Engineering we could suppose the following research grounded on the positivist stance: use of controlled studies (methodology *M*) to investigate whether software maintainability (area of concern *A*) can be improved by object orientation (conceptual framework *F*). See top of

Figure 3.

However, Action Research changes the role of *F*, *M*, and even *A*, as the researcher becomes involved in the transformations occurring in a given situation (Checkland and Holwell, 1998). This way, the researcher interested in a research theme can declare *F* and *M* and then go into a real-world situation related to the area of concern *A* in which that theme is relevant. As a result, Action Research not only represents *M* as in the Natural Sciences, but requires a research protocol merged into *M*, *which* is embedded with the practical and scientific conceptualizations from *F* allowing the research process to be recoverable by anyone interested in submitting the research to critical analysis (see the bottom of that without it the evaluation of the results tends to be compromised, and eventually what is being described may be action (but not research) or research (but not action).

Figure 3). Nevertheless, even with this, Action Research does not obtain the repeatability property present in natural sciences, but it is sufficient to clarify the processes and models that allowed the researcher to make the interpretations and derive the conclusions. The importance of making the criteria explicit before conducting the research is also emphasized by Avison *et al*. (1999) who claim

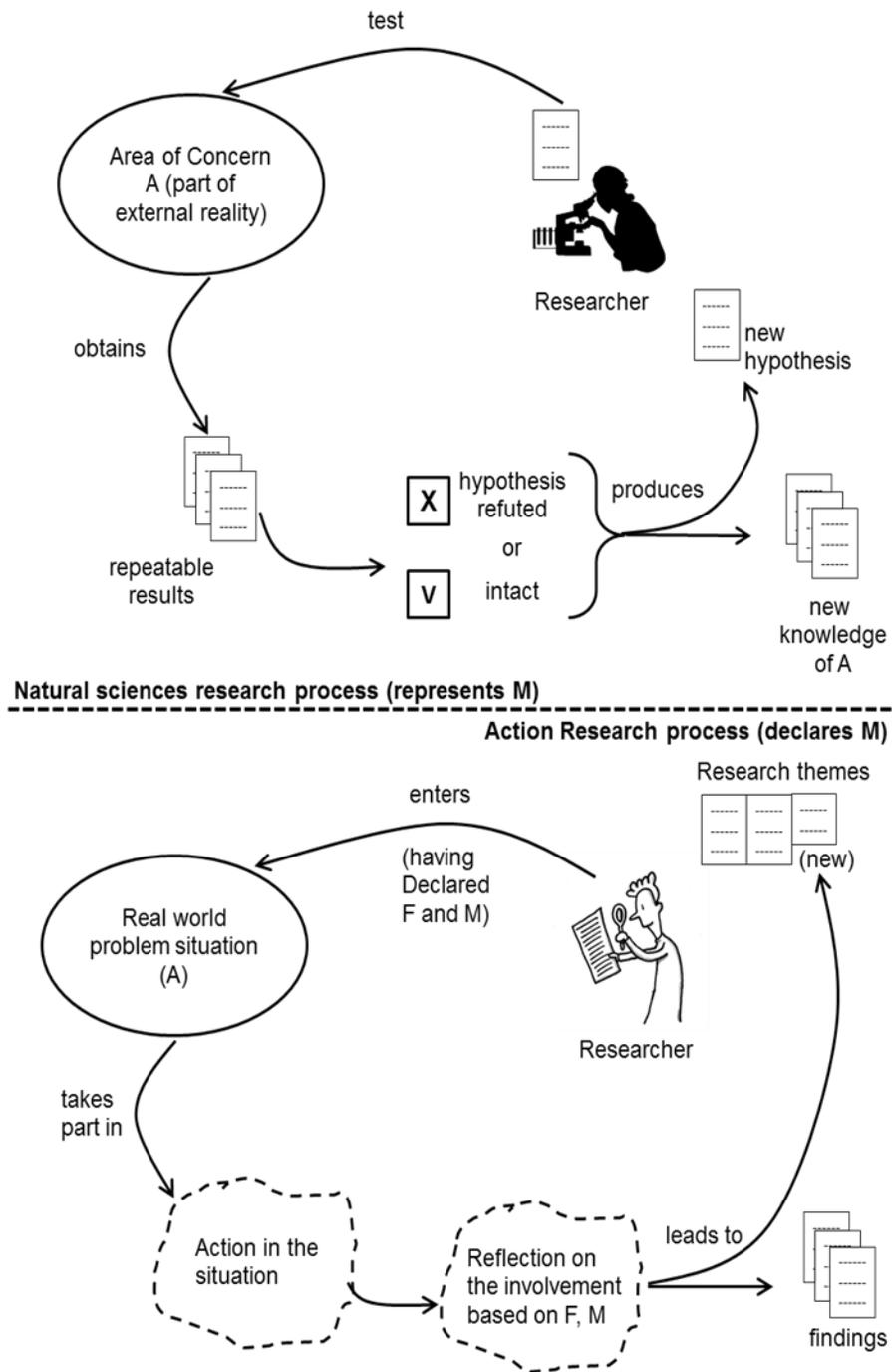

Figure 3 – Action research and Natural Sciences - process comparison – adapted from Checkland and Holwell (1998)

The process shown in

Figure 3 provides important indications of issues that should be addressed to provide rigour to Action Research but it still does not have a detailed guide explaining how the research should be conducted. For this purpose, other researchers have expanded the basic framework that guides the process. The most common process is from Susman and Evered (1978). According to Davison et al. (2004), it has achieved the status of 'canonical' process, consisting of five stages (Figure 4):

(1) **Diagnostic:** consists of exploring the research field, stakeholders and their expectations holistically. In this stage, there is also the research theme definition that is represented by the designation of the practical problem and knowledge area to be addressed.

(2) **Planning:** stage where actions are defined to the faced circumstances. These definitions are guided by hypotheses portraying the researchers' formulated assumptions about possible solutions and results. These hypotheses, on the other hand, should follow the scientific theoretical formulation.

(3) **Intervention:** corresponds to the planned actions implementation. An essential element in this stage is the seminar technique that can be used to exam, discuss and make decisions about the investigation process (Thiollent, 2007).

(4) **Evaluation:** stage where the interventions effect are analyzed considering the theoretical background used as basis to the actions definition.

(5) **Reflection:** involves the dissemination of acquired knowledge among participants and other organization departments. The learning experience is facilitated by the previous collaboration among participants and researchers in the technical topics.

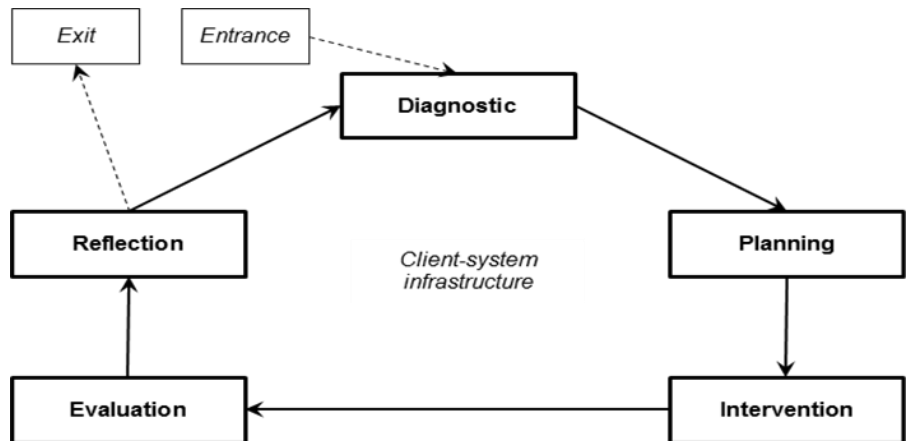
**Figure 4 – Action Research canonical process – based on Davison *et al.* (2004)**

Apart from these stages, the research environment requires a contract/agreement to legitimize the actions and possible benefits for both parties (researchers and organization), which builds up the so-called client-system infrastructure. There are two additional steps that are not directly part of the cycle but relate to the start and end of the Action Research process. For Avison *et al*. (2001), the start of an action research study requires attention as there should exist a perceived need for real improvements by the client, which has to demand scientific support. Otherwise there is a risk of one starting studies focused on irrelevant themes (where there is no perspective to the researcher for generating knowledge) or not starting studies as the so-called 'iceberg' themes (where the client does not discern the need for improvements and therefore does not seek help from a researcher). The other additional step regards the criteria to finish the study. This should occur whenever both parties involved are satisfied with the outcome. That is why a contract/agreement is so important, to clearly state interests such as this. While this condition is not achieved, the Action Research process allows iterations to enable the achievement of results incrementally. This iterative process is especially advantageous when the initial diagnosis cannot be entirely performed.

Moreover, the cycle defined in

Figure 4, although not explicitly represented, allows some iteration and adaptation between the stages. For instance, the researcher can return to the diagnosis stage after an initial attempt to plan the study has not been completed because of a lack of better problem description, or if the researcher

can update the study plan during the course of action taking some unexpected events into consideration.

### 2.2.1. Other Action Research Characteristics

The process described in the previous section already anticipated some of the key characteristics of Action Research such as a cyclical process model, the need for a agreement/contract between the client and the researcher, and learning through reflection. Additionally, Davison *et al*. (2004) define two essential Action Research characteristics that the authors call 'principles of canonical action research':

- **Theory principle:** researchers should ground themselves upon theories as a way to guide and focus their activities. Theories help not only conducting the research and taking actions to solve a problem, but also support on reporting study results and positioning them in the existing accomplished research in the field. In most cases theories are presented this way: in a situation *S* with evident environmental conditions *F*, *G*, and *H*, results *X*, *Y*, and *Z* are expected from actions *A*, *B*, and *C*.

- **Change through the action principle:** defines the essence of Action Research which is the indivisibility between research and action. The lack of action in a problem situation suggests the absence of a meaningful problem. Actions have to be planned in order to address the observed problems allowing the researcher to justify each action or reparation of part or the whole of the problem diagnosed.

All these characteristics make Action Research singular amongst the research methodologies. Action Research differentiates itself from routine practice because it is proactive and strategically driven by theories and research techniques. At the same time, it distinguishes itself from 'normal' scientific research because of its flexibility in recognizing the importance of collaboration between researchers and subjects, and the value of using the best available evidence even when not having a good baseline to make precise judgments on the outcomes (Tripp, 2005).

## 3. The Use of Action Research in Software Engineering

To investigate the use of Action Research in Software Engineering a technical literature survey has been conducted. We selected nine major Software Engineering journals and three conference proceedings in the period of 1993 to June 2010. The results up to June 2009 were presented in (Santos and Travassos, 2009). In this section, we provide the updated results up to June 2010. As it will be presented later, these new results reinforce the increasing interest in the Action Research methodology by the Software Engineering community.

There are other technical literature surveys in Software Engineering that investigated particular categories of empirical studies including controlled experiments (Sjøberg *et al.*, 2005) and quasi-experiments (Kampenes *et al.*, 2009), but as far as these authors are aware there is no one specifically concerned with Action Research. We identified 22 papers in the journals and conferences – six found in the last update (July 2009 to June 2010). To update the results we ran the same search strings and used the same information extraction form as described in (Santos and Travassos, 2009). Although Action Research studies represent a very small fraction of the studies being conducted in Software Engineering, they are concerned with different Software Engineering contexts and thus are sufficient for exemplify to researchers the potentials of Action Research. However, the results of the survey have shown a better definition of what can be considered an Action Research study in Software Engineering is needed. For instance, several studies that self reported to be Action Research were in fact case studies (and vice-versa). The overloading in the classification of empirical studies was also previously observed by Sjøberg *et al.* (2007).

## 3.1. Method

To conduct this survey, we followed some of the criteria and steps of the approach employed in other technical literature reviews (Lau, 1997, Sjøberg *et al.*, 2005). The journal and conferences chosen are the same ones as Sjøberg *et al.* (2005), which are considered relevant to Software Engineering research. The journals are the ACM Transactions on Software Engineering Methodology (TOSEM), Empirical Software Engineering (EMSE), IEEE Computer, IEEE Software, IEEE Transactions on Software Engineering (TSE), Information and Software Technology (IST), Journal of Systems and Software (JSS), Software Maintenance and Evolution (SME), and Software: Practice and Experience (SP&E). The conferences are the International Conference on Software Engineering (ICSE), the IEEE International Symposium on Empirical Software Engineering (ISESE), and the IEEE International Symposium on Software Metrics (METRICS). The International Symposium on Empirical Software Engineering and Measurement (ESEM) was included in this survey as it merged the METRICS and ISESE conferences used in Sjøberg *et al.* (2005) beginning in 2007.

The terms used for the search were gathered from Baskerville and Wood-Harper (1998): *action research, action learning, action science, reflective practice, critical systems theory, systems thinking, and participative research*. Using these terms we found 189 papers from which 22 were selected. From the initial 189 papers, 151 were eliminated by title and abstract, and the remaining 31 were entirely read. The main applied criterion when selecting the papers was to evaluate whether it really represented an Action Research study. To do this we drove our decision based on the set of criteria for acceptable Action Research studies given by Lau (1997) (mentioned in the previous section): a real need for change, theory-based iterative problem solving, genuine collaboration with participants and honesty in theorizing research from reflection; amended by the Action Research principles defined by Davison *et al.* (2004): researcher-client agreement, cyclical process model, theory use, change through action and learning through reflection. However, driven by the presupposition that the finding of an Action Research study in Software Engineering that met all these criteria and principles would be difficult, we defined Action Research adherence levels to classify the studies. It aims at being more open when selecting the papers. The idea to have the adherence levels came from one of the selected papers where the authors explicitly stated that they conducted the research study inspired by the Action Research methodology (Abrahamsson and Koskela, 2004). All the information extracted from the papers is summarized in

Table 1.

**Table 1 – Information extracted from the papers**

| Information | Description | Based on |
|---|---|---|
| Problem | The research problem, usually related to the AR diagnosis. | (Lau, 1997) |
| Action | Action implemented. | (Lau, 1997) |
| Reflection | Reflections from the actions implemented and problem solution. | (Lau, 1997) |
| IEEE Taxonomy | Used to classify the research topics. | (IEEE Keyword Taxonomy, 2002) |
| Adherence | <u>Inspired</u> – when the focus is on the researchers learning from a real problem resolution exploring SE research without controlling the study by the AR principles;<br><u>Based</u> – when the AR methodology is modified or combined with other empirical methods;<br><u>Genuine</u> – when the full essence of action research methodology is present. | – |
| Type | <u>Action Research</u> – focusing on change and reflection;<br><u>Action Science</u> – trying to solve conflicts between espoused and applied | (Avison et al., 1999) |

| Information | Description | | Based on |
|---|---|---|---|
| | theories; Participatory Action Research – emphasizing participant collaboration; Action Learning – for programmed instruction and experiential learning. | | |
| Length | Length of the study. | | (Sjøberg *et al.*, 2005) |
| Data Collection | Qualitative or Quantitative, including the techniques used for data collection. | | - |
| AR control structures | Initialization | Researcher – field experiment; Practitioner – classic action research genesis; Collaboration – evolves from existing interaction. | (Avison *et al.*, 2001) |
| | Authority | Practitioner – consultative action warrant; Staged – migration of power to the client; Identity – practitioner and researcher are the same. | |
| | Formalization | Formal – specific written contract; Informal – broad, perhaps verbal, agreements; Evolved – informal or formal shift into opposite form. | |
| AR cycles | Number of AR cycles conducted. | | (Davison *et al.*, 2004) |

We tried to be rigorous by selecting only SE research papers, as some journals and conferences also contain Information Systems items. SE applies Computer Science fundamentals to the development of software systems. Information Systems are concerned with the business community needs in terms of computing, and especially information. Thus both SE and Information Systems fields have certain elements in common – computing concepts, systems development, and information technology – but they also have clearly distinguishable goals (Ramesh *et al.*, 2004).

### 3.2. Results

In this section we present three major results. First, the distribution of the publications along the years, journals and conferences are shown. Then we have two subsections to characterize the domains and contexts within which action research studies were conducted and that describe how they were done.

Publication distribution is shown in Table 2 (only journals and conferences from where papers were selected are listed). Adherence levels defined earlier in Table 1 are now abbreviated in Table 2 as Inspired (*I*), Based (*B*), and Genuine (*G*). From the distribution it is possible to see that the number of reports on Action Research studies presents a smooth increase along the years, with more studies being reported in the last two years. Notice that in the last period (2005 - June 2010) of data collection some 2010 conferences and journals had not yet published their proceedings (e.g., ESEM), thus providing a partial view of this period.

Table 2 – Distribution of the Selected Papers

| Journals and Conf. | 1993-1998 | | | 1999-2004 | | | 2005-2010 | | |
|---|---|---|---|---|---|---|---|---|---|
| | (I) | (B) | (G) | (I) | (B) | (G) | (I) | (B) | (G) |
| ESE | | | | 1 | | | 1 | | |
| ICSE | | | | | | 2 | | | 2 |
| IEEE SW | | | | 1 | | | 2 | | |
| IST | | | | | 1 | 1 | | | 5 |
| SME | | | | | | | | 1 | |
| SP&E | 1 | | | | | 1 | 1 | | |
| TSE | | | | | | | | 1 | 1 |
| Totals | 1 | | | 2 | 1 | 4 | 3 | 3 | 8 |
| | 1 | | | 7 | | | 14 | | |

The number of studies inspired by Action Research is relatively high, about 30%. This means that there is a need to improve rigour in Action Research studies as to whether we want Action Research investigations to form a solid ground for further research and industrial applications in SE. This situation is even worse if we consider the studies that mentioned Action Research but were eliminated as they could not even be classified in the lowest (i.e., inspired) adherence level.

### 3.2.1. Research Topics and Contexts

It is interesting to see that Action Research is being applied to a wide spectrum of Software Engineering research domains (Table 3), ranging from the more social side (e.g., Management and Software Engineering Process) to the more technical end (e.g., Software Construction and Programming Environments). The topic with the wider number of studies is Software Engineering Process, more specifically Process Implementation and Change. This is also the topic where most of the inspired adherence level articles were concentrated, mainly because their authors interleaved the software improvement process with the Action Research process leaving implicit when they were assuming the researcher or the consulting roles. Other topics include architecture knowledge management, agile methods, component-based development, scientific workflows, software inspection, and security.

Table 3 - Research Topics According to the IEEE Taxonomy

| IEEE Taxonomy | # Articles |
|---|---|
| **Distribution, Maintenance, and Enhancement** | 2 |
|     Documentation | (Lindvall and Sandahl, 1996) |
|     Maintenance management | (Polo et al., 2002) |
| **Management** | 3 |
|     Project control & modelling | (Canfora et al., 2006) |
|     Time Estimation | (Staron and Meding, 2008) |
|     Risk Management | (McCaffery et al., 2009) |
| **Programming Environments/Construction Tools** | 1 |
|     Environments for multiple-processor systems | (Vigder et al., 2008) |
| **Requirements/Specifications** | 3 |
|     Elicitation methods | (Maiden and Robertson, 2005) |
| | (Napier et al., 2009) |
|     Process | (Kauppinen et al., 2004) |
| **Reusable Software** | 1 |
|     Reuse Models | (Bosch, 2010) |
| **Software and System Safety** | 1 |
|     Software and System Safety | (Gutierrez et al., 2009) |
| **Software Architectures** | 2 |
|     Domain-specific architectures | (Bengtsson and Bosch, 1999) |
|     Patterns | (Mattsson et al., 2009) |
| **Software Construction** | 2 |
|     Data design and management | (Fernández-Medina and Piattini, 2005) |
|     Programming paradigms | (Lycett, 2001) |
| **Software Engineering Process** | 6 |
|     Process implementation and change | (Fitzgerald and O'Kane, 1999) |
| | (Kautz et al., 2000) |
| | (Salo and Abrahamsson, 2005) |
| | (Nielson and Tjørnehøj, 2010) |
| | (Pino et al., 2010) |
|     Software Process Models | (Abrahamsson and Koskela, 2004) |
| **Miscellaneous** | 1 |
|     Software libraries | (Staron et al., 2009) |

As regards the diagnosis/action/reflection extracted information we could classify it into three major formats: (1) evaluation of technology introduction through lessons learned (presenting similarities with case studies), (2) technology conception and/or tailoring with intense collaboration and changing through intervention focus (the AR most genuine format where problem solution is initially unknown), and (3) SE activities facilitation and observation (having a consulting component).

In general, the more technical initiatives were closely related to formats (1) and (2). For example, one of the papers reported a maintenance methodology created in the context of an organization, while in another case the formalization of software architecture design rules in the context of model-driven development was introduced. On the other hand, the more social research efforts were more related to the format (3) as it was the case of the software process improvement papers.

**3.2.2. Execution Details**

Many papers did not describe the data collection techniques, study length, and number of Action Research cycles. Therefore, the extraction of information on execution details was problematic. Moreover, almost all papers did not explicitly define the Action Research type and control structures. The Action Research control structures are an important component of an Action Research study execution as it reveals the process followed, and why decisions were taken. The definition of the Action Research type, in its turn, shows why some Action Research characteristics were emphasized. Although this information was not explicit, we could in some cases implicitly deduce them from the overall research actions and context description. Figure 5 shows the number of papers as per Action Research type and control structures. In examining Figure 5 we can see that most of the reported studies described in the papers are of the Action Research type, and are started by practitioners having authority over the research execution (identity authority) that is carried out without any formalization. This appears to be an interesting find as it possibly means that researchers are conducting studies considering a small number of organizational constraints. Nevertheless, it is worth reiterating that a large percentage of the papers do not mention their control structures.

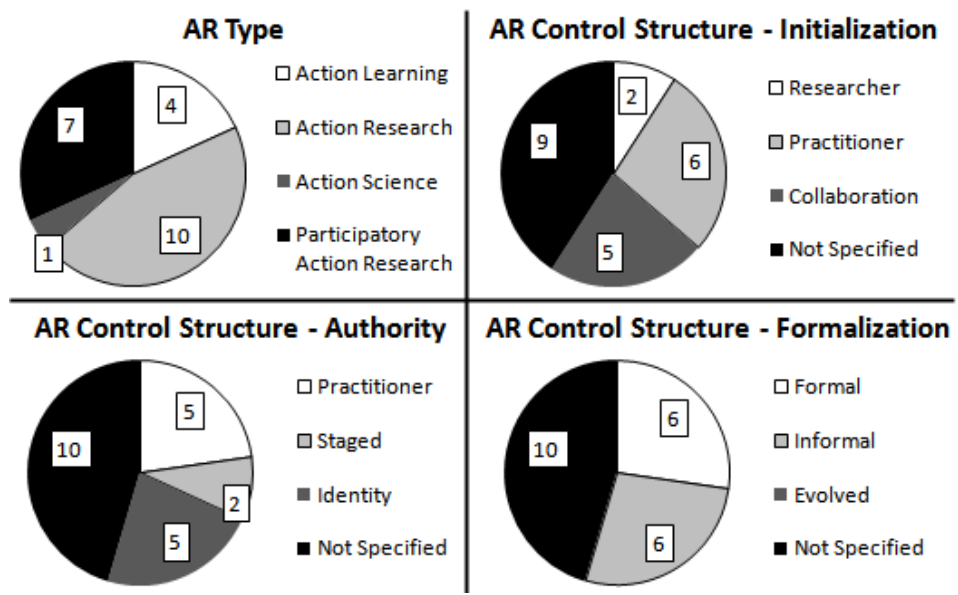

Figure 5 - Classification of selected papers as per Action Research types and control structures defined in

**Table 1**

All papers made intensive use of qualitative data, confirming this intrinsic characteristic of Action Research in Software Engineering. Four studies used quantitative data, indicating that quantitative research is also a possibility in Action Research. Observation was by far the most mentioned data collection technique, followed closely by interviews. For quantitative data, metrics were reported to be used in all papers.

Finally, the length of the studies ranged from 2 months to 5 years (mean time was 21 months and 16 for the standard deviation). This result shows that the use of Action Research is very flexible regarding the study duration and is most influenced by the research topic, software technology, and activities involved. Five studies did not specify their length. For the number of action research cycles, only three papers explicitly mentioned to be 1 cycle. But, from the linear description in the other papers we believe the same behaviour can be expected for most of the other cases.

## 4. Using Action Research in Software Engineering: an *in vivo* study

The research theme addressed in this study concerns source code refactoring. Two very distinct aspects of this theme are investigated. The initial objective when starting this study was to use the software refactoring process to externalize knowledge about architectural styles and source code conventions in the software project. However, due to the features of Action Research the researchers were motivated to understand how software developers decide the appropriate refactoring moment regarding some piece of source code.

The organization of the study documentation has been derived during its execution. From this experience, we would like to suggest a template to be used in other Action Research studies in Software Engineering. The template structure is based on the Action Research stages consisting of one section for each stage, including some additional subsections intending to better explain the whole study. Details on how the template can be used to support the reporting of Action Research studies are shown in Section 5.2.3.

Even though the results of the study were relevant to the software project, this section is intended to illustrate how an Action Research study could be conducted and documented. Therefore, we would like to remind the readers to focus on these aspects in this section.

## 4.1. Diagnosis

### 4.1.1. Problem Description

Architectural styles and coding conventions represent essential elements in the construction stage of software development, associated with important quality characteristics such as readability and maintainability. In addition, these elements have strong social and managerial implications as they can support the project team organization (Kazman and Bass, 2002).

The definition and documentation of architectural styles and coding conventions should be ***a priori*** worked on in a regular basis. However, in teams with a limited number of members these issues can be worked on by means of direct communication. On the other hand, when new members join the team, including remote ones, direct communication turns out to be not the appropriate medium for this kind of knowledge dissemination, especially considering when communication can compromise the productivity of the developers and affect software quality. Additionally, there are two issues that should be considered into this context: (1) new developers learning the process overloads communication, and (2) remote teams introduce difficulties to the communication amongst all the members of the team.

To reduce the impact on the communication of knowledge amongst team members (being previously done through direct dissemination amongst the local ones) knowledge should be externalized and explicitly described. One immediate way to proceed in knowledge externalization could be to use the more experienced software project developers to explicit it. However, due to their larger experience, there is a possibility that more experienced developers would not be able to understand the real needs of the less experienced members in terms of information, as these basic needs are not part of the more experienced developers daily activities.

Presented with these risks, this study investigated mechanisms to allow knowledge externalization related to architectural styles and coding conventions, considering the needs of new project team members, mainly the remote ones.

### 4.1.2. Project Context

The software project within which the research was conducted aims at the developing of a new Web-based information system to support the activities concerned with the management of R&D engineering projects in the context of a non-profit Science and Technology Foundation (STF) at the Federal University of Rio de Janeiro. It represents a medium-to-large software project including seven modules (over two hundred use cases) and covering different organization departments such as human resources, accounting, and project management. The software was modularized and developed following an incremental life cycle. Amongst the factors that justified this decision we can mention the stakeholder interest (1) on partial deliveries of the product, and (2) on the gradual replacement of the current system by the new one. The STF follows a reference software development process model based on MPS.BR from the G up to the C level (SOFTEX, 2007). MPS.BR is a nationwide program for software process improvement in Brazilian organizations. The main goal of this initiative is to develop and disseminate a Brazilian software process model (named MPS Model) aimed at establishing a feasible path for organizations to achieve benefits from implementing software process improvement at reasonable costs, especially in small and mid-size enterprises. The performance of companies using the MPS-BR has been followed via periodical assessments, and results are positive (Travassos and Kalinowski, 2009).

The STF new information system was developed using Java technology. The designed software architecture is based on the layered model-view-controller (MVC) architectural style where the View layer was implemented using JavaServer Faces (JSF – a Web framework that uses graphic interface components to create Web interfaces) and the Model and Controller layers using plain Java. At the time of the study (November 2008), the team was distributed and consisted of three software designers and two software developers at the STF site (denominated by team L) plus six developers physically located in another city 200km away from Rio de Janeiro (team R). The average experience of the team L with software development was of about seven years (varying between 3 and 15 years), whilst the team R had about 2.5 years (varying between 1.5 and 4 years). Moreover, the team L had three professionals responsible for specification and quality assurance activities.

Each module construction was implemented in iterations made by a set of conceptually related use cases. In each iteration the software design stage consisted of the designers defining business classes to form the Model layer. After the design was released, the developers at the remote site were responsible to implement both the View and Controller layers. The communication between the teams used Internet apparatuses such as email, instant messaging systems, and video conference. It is also worth mentioning that all software project issues such as defects, adjustments, and tasks were registered and monitored using the TRAC system (issue tracking system) (Trac, 2003).

The detailed documentation on the architectural style and coding conventions used was not an outcome of the project development process. Most of these definitions were brought from the Sun official documentation (Sun, 1999) and industry recommendations (Crupi *et al.*, 2004). However, some adaptations and modifications were made by the development team. One of them, for example, regards the creation of an own mechanism to perform the validation of input data forms, ignoring the ones provided by the JSF technology.

In March 2008, time when a new software project module development was starting, Team R was added to the project. At that time, this represented an increase from eight to fourteen professionals. These new Team R members received on-site face-to-face training for two days. During the training sessions the technology used in the project was presented and issues regarding the construction procedures were discussed. The following topics were addressed:

- The Java and JavaServer Faces technologies;
- The model-view-controller architectural style, including discussions on object-oriented paradigm concepts, such as encapsulation and reuse through inheritance;
- Source code naming conventions and organization. This was presented to re-implement a previous use case from an earlier software project module;
- Configuration management activities; and
- The development process overview, including the specification and testing activities.

In spite of these initial training and monitoring activities having been conducted, the more experienced members from team L some identified some issues related to the quality of the source code as implemented at the end of the first iteration of the members of team R. Even though this behaviour is not desired it is common to observe it in other projects due to, amongst other reasons, the lack of familiarity with the project (problem domain) even if the technologies used are well known (Bennet and Rajlich, 2000). These issues regarded the failure to follow source code organization according to tacit coding conventions and architectural styles expected by the more experienced members.

### 4.1.3. Research Theme

The research theme concerned with the previously presented scenario can be defined as follows: *strategies to detect non-conformances in source code caused by lack of knowledge of the software project coding conventions and architectural styles*. The intention is to use refactoring techniques to identify coding non-conformances when compared with the tacit knowledge held by the experienced developers. This way, we worked on source code refactoring aiming at improving its quality as usually perceived by the more experienced developers in the software project.

According to Mens and Tourwe (2004), the main idea in refactoring is to redistribute or rewrite classes, attributes, and methods through classes' hierarchy, to facilitate further adaptation and extension. In other words, the goal in refactoring is to improve the software's internal structure, keeping its external behaviour. Thus, based on the non-conformances detected in the less experienced developers' implemented code, we expected that it could be possible to define what knowledge on source code conventions should become explicit. To support this expectation the fact that inspections/reviews are a useful tool for organizational learning is being explored, as good development patterns can be observed during the reading for refactoring.

## 4.2. Planning

In this section the planning stage is described. It begins with a technical literature survey where some works on the research theme can be examined. Based on that, the intervention focus can be set and the hypotheses associated to the outcomes can be defined. Moreover, the instruments, tools, and techniques that are expected to be used in the research are presented.

### 4.2.1. Literature Survey

Given the research theme the search for published papers in the technical literature focused on inspection techniques that could offer some guidance on how the source code should be read and could also assist in revealing non-conformances. We sought techniques requiring as little training effort as possible. Just the explanation on their application should be enough to allow their application. Additionally, we looked for technical articles introducing mechanisms that could reveal the underlying reasons for the captured non-conformances. The literature survey was conducted on an ***ad-hoc*** basis, and papers were selected by convenience.

Two studies were selected in this survey: (Dunsmore et al., 2003) and (Mäntylä, 2005). They set the ground to design the procedure for source code reading (see form in Table 4). Dunsmore *et al.* (2003) describe a use case based inspection technique that allows object-oriented source code reading by exploring its dynamic model perspective. The steps of the technique encompass (1) *the selection of one use case*, (2) *the derivation of scenarios from it (e.g., 'save invoice', 'cancel invoice')*, and (3) *reading the class methods responsible for scenario execution*.

In order to categorize the refactoring suggestions produced by the source code inspection we used the concepts from a Mäntylä and Lassenius (2006) study which is subsequent to the Mäntylä (2005) study and directly related to it. Mäntylä and Lassenius (2006) describe an experimental study aimed at understanding why and when developers perceive the need for source code refactoring. It is a qualitative study where subjects were instructed to register why they felt the source code should undergo refactoring. The authors analyzed these records, extracting a taxonomy of defects (poor algorithm, poor internal organization, minor structure issues, duplicated code, too many temporary variables, poor readability, poor method name, wrong indentation, poor comments, poor parameter layout, long method, or extracted method). We decided to explore the same taxonomy during our source code refactoring process. To do that we included the following step into the procedure described in Table 4: (4) *registering refactoring decisions for each method according to the previous knowledge of the developers*. Moreover, Mäntylä (2005) suggests a set of source code metrics that could be used to evaluate some properties of the source code under refactoring. These metrics were also used in our study aiming at comparing the results of the studies and will be described in the next section.

### 4.2.2. Action Focus
*4.2.2.1. Objectives*

We defined the goals of the study using the Goal-Question-Metric (GQM) approach as proposed by Basili *et al*. (1994), as follows:

> *analyze* the using of source code refactoring *for the purpose of* characterizing *with respect to* externalization of knowledge associated with the architectural styles and coding conventions *from the point-of-view* of the software engineers *in the context of the development of W*eb information systems by distributed teams.

Additionally, a secondary goal was included to compare the results of this study with Mäntylä (2005):

*analyze* source code measurable characteristics *for the purpose of* characterizing (can it explain the refactoring decision subjective evaluation?) *with respect to* the predictability of the measurements *from the point-of-view* of the software engineers *in the context of the development of W*eb information systems by distributed teams.

*4.2.2.2. Research Questions*

The objectives will be achieved when answers (results) have been given to the following questions (in reverse order of objectives):

- Q.1. What source code aspects or characteristics does a developer use when identifying refactoring as necessary? This question aims at understanding the primary motivation to experienced developers identify a refactoring opportunity.

- Q.2. Is source code refactoring useful to externalize knowledge on architectural styles and coding conventions used in a software project, particularly those a developer needs when getting involved in new projects? The idea of this question is to assess whether the refactoring opportunities identified in question 1 can be used as an input in documenting the architectural styles and coding conventions used in the project.

*4.2.2.3. Expected Outcomes (data collection)*

Question Q.1 has the following operational questions associated:

- Q.1.1. What is the refactoring effect over the source code metrics?
- Q.1.2. What are the source code characteristics that affect developer decisions?
- Q.1.3. What types of refactoring are exclusively identified by the experienced developers?
- Q.1.4. Is it possible to classify the identified refactoring opportunities to reflect expected source code quality characteristics?
- Q.1.5. What is the perception of the developers on the refactoring effect over source code quality?

The source code metrics suggested in Mäntylä (2005) (Lines of Code, Number of Parameters, Cyclomatic Complexity, Invoked Methods, Fan-in and Fan-out) will be used to support answers to questions Q1.1 and Q1.2. According to Mäntylä (2005) these metrics were selected based on their acceptance described in the technical literature, aiming at characterizing the methods from different perspectives. A brief description for each metric is given:

- <u>Lines of Code</u>: number of lines in source code (including comments).
- <u>Number of Parameters</u>: number of parameters of a method.
- <u>Cyclomatic Complexity</u>: #edges - #vertices + #connected components in a control flow graph.
- <u>Invoked methods</u>**:** number of methods invoked inside a method.
- <u>Fan-in</u>: number of input variables used in a method, including its parameters and global variables.
- <u>Fan-out</u>: number of output variables updated in a method, including the return command and global variables.

Returning to the operational questions, we planned a categorization activity with the developers based on the Mäntylä and Lassenius (2006) taxonomy of defects to answer Question Q.1.4. Finally, questions Q.1.3 and Q.1.5 will be answered in interviews with the developers.

Question Q.2 is associated to the following operational questions:

- Q.2.1 Should the description of the refactoring opportunities as identified by the developers contain details on what motivated their identification?
- Q.2.2 Should the source code refactoring be performed by those who wrote the source code in order to facilitate their grasping the tacit knowledge involved in the coding conventions?
- Q.2.3 How can tacit knowledge be acquired and made explicit?

Question Q.2.1 will be captured by the form presented on Table 4. In its turn, Question Q2.2 just requires project management activities to be answered. Last but not least, Question Q.2.3 will be discussed between the project members from both teams (L and R).

### 4.2.3. Hypotheses (suppositions)

- There is some tacit knowledge concerned with the project organization conforming with an architectural style, and with coding conventions.

- The use of the Dunsmore *et al.* (2003) inspection technique of the source code refactoring will allow knowledge externalization related to the architectural styles and coding conventions used in the software project.

### 4.2.4. Operational Definitions

*4.2.4.1. Techniques*

Semi-structured interviews will be the main qualitative data collection approach during the interventions. Additionally, characterization forms will also be filled in by the participants. Quantitative data will be directly extracted from the source code and analyzed using descriptive and predictive statistics.

*4.2.4.2. Instruments*

Table 4 shows the information contained in the discrepancy (non-conformance or refactoring suggestion) form that will be used by the developers to describe the source code refactoring opportunities. Apart from the fields that should be filled by the developers, the form has the description of the procedure steps that should be observed to facilitate source code reading and understanding. According to Dunsmore *et al.* (2003), source code comprehension can be simplified as it forces the reading of one use case in its turn. Steps 1, 2, and 3 were directly brought from Dunsmore *et al.* (2003) while fields 5 and 6 were brought from Mäntylä and Lassenius (2006). The other fields and steps were formulated specifically for the context of this study.

Table 4 - Form content used in the study

| Procedure steps |
|---|
| 1 - Select one use case from the software |
| 2 - Extract use scenarios from the use case (e.g., 'save invoice', 'cancel invoice'); |
| 3 - Read through the class methods responsible for scenario execution. Check if the correct methods are being invoked and if the scenario state is being kept and manipulated in a consistent way by the system; |
| 4 - For each class method, register the refactoring opportunities according to your previous knowledge of the project (e.g., excessive nesting, naming, visual source code organization, amongst others). |
| **Form fields** |
| 1. Use case |
| 2. Scenario description |
| 3. Class name |
| 4. Method name |
| 5. Would you apply refactoring to this method? (answer from 1 to 5: 1 – No; 2 – Unlikely; 3 – Likely; 4 – Yes, but only at another moment in the project; 5 – Yes, right now) |

> 6. Explain your decision to the above question. If refactoring is needed explain why, what, and how the method should be improved. If the method is OK explain what quality attributes are being met, including agreed project patterns. If your answer is 'Likely' provide your reasoning too. If appropriate, mention the source code lines involved in your explanation.

Two interviews were planned aiming at discussing the topics shown on Table 5 and Table 6.

Table 5 - Semi-structured interview 1

| Refactoring related topics |  |
|---|---|
| (i) | Were the questions and procedure suggested useful to focus the refactoring process on the agreed purpose? |
| (ii) | How would you rate each given explanation for the refactoring opportunities using the Mäntylä e Lassenius (2006) taxonomy of defects? (Q.1.4) |
| (iii) | What is the refactoring effect on source code quality? (Q.1.5) |
| (iv) | What kind of refactoring do you think could only be identified by the experienced developers? (Q.1.3) |
| (v) | Did you miss any characteristic to be captured on the given form? Were the blank fields enough for the exceptions? |

Table 6 - Semi-structured interview 2

| Knowledge externalization topics |  |
|---|---|
| (vi) | How did the identified refactoring opportunities facilitate the learning of coding conventions and architectural styles as used in the project? |
| (vii) | Decision on how to formalize the externalized knowledge. (Q.2.3) |
| (viii) | In your opinion was the refactoring useful as a means to externalizing knowledge? Do you still feel that some knowledge was not externalized? |

*4.2.4.3. Tools*

For the extraction of source code metrics and the quantitative data analysis the tools 'Understand – Source Code Analysis & Metrics' (http://www.scitools.com/) and SPSS (http://www.spss.com/) tools, respectively, were used.

## 4.3. Actions

Initially, development team R was notified on some issues regarding quality as seen in the source code. Most of them were due to communication failures and the learning process of working in a distributed way. For this reason, source code refactoring should be started trying to show the team how the source code should be structured. Two more experienced developers from team L (a designer/developer, and a developer) were selected to perform the source code review and identify the refactoring opportunities. Each one of them was in charge of reviewing different use cases.

The procedure steps and form fields (Table 4) were presented to the developers who also received instructions on how to fill them in. The instructions basically included the review focus, which gave overall instructions on how to search for source code non-conformances according to what developers perceived as coding conventions or architectural styles tacitly used in the project. The definition of non-conformance was kept generic in an attempt to maintain the subjectivity in refactoring decisions and to capture richer descriptions of the opportunities. Some examples were discussed using the prior knowledge of developers on refactoring, including concepts on the object-oriented paradigm (inheritance, readability, maintainability, amongst others). Following the instructions, the refactoring process was started and after its completion the information filled in the forms was sent to the developers who originally built the use cases.

Based on the refactoring opportunities identified and registered in the forms by team L members, team R was instructed to apply refactoring to the source code. It was also instructed that the original source code versions should be recorded (using the version control system) together with the time spent with each method that received refactoring. As planned, those responsible for the construction of the use cases applied refactoring to their own source code, but could disagree on the suggested refactoring opportunities.

Lastly, we started the planned semi-structured interviews (Tables 5 and 6). Interviews 1 and 2 were made with the developers. The interviews were first recorded and next could be transcribed. Besides, after the interview, team L developers were asked to rate their refactoring opportunities with the researcher using the Mäntylä and Lassenius (2006) taxonomy for defect categories.

### 4.4. Evaluation and Analysis

In this section two different types of data will be analyzed. One is the source code metrics extracted before and after refactoring, in an attempt to answer questions Q.1.1, Q.1.2, and Q.1.4. The other is concerned with the interviews transcript, used to discuss questions Q.1.3, Q.1.5 and Q.2.3.

#### 4.4.1. Analyzing Source Code Metrics

As previously discussed, each developer described what motivated the identification of each suggested refactoring opportunity. Based on this description, the defect taxonomy was applied and evolved. The categories of defects (see section 4.2.1) as well as a summary of the refactoring effects on the source code metrics considering each defect category are shown in Table 7. Two new defect categories have emerged in the context of this study - 'low cohesion' and 'unused code' (underlined in Table 7). The developers reviewed 47 methods. In 37 of them there was at least one refactoring opportunity. The average size of the methods was 32 lines of code (with large variation, σ = 34).

To observe the effect on the metrics it is interesting to split the refactoring suggestions into two categories: defects and improvements. Both types are separated in Table 7 by an empty line. In each row (Table 7) the values were calculated by subtracting the metric value before and after the refactoring and summed up by category. Thus, negative values represent a decrease in the metric value after the refactoring. Even when the developers were not asked, they found additional defects during the review. Many of these defects, based on their descriptions, were related to incomplete features or, in other words, omission defects. As a result, we can see through the metrics that the correction of defects increased both the source code size and complexity. On the other hand, analyzing the improvement category for refactoring opportunities we can see (Table 7 Total line) that the result was opposite to the former. In this case, applying refactoring meant a reduction in source code size, complexity, and coupling (considering the fan-in, fan-out, and remote methods' invoked metrics).

Table 7 – Metrics variation before and after source code refactoring

| Category | Refactoring opportunities | Cyclomatic Complexity | Fan-In | Fan-Out | Lines of Code | Remote Methods Invoked |
|---|---|---|---|---|---|---|
| Defect | 6 | 5 | 5 | 7 | 60 | 7 |
| | | | | | | |
| Poor Algorithm | 7 (17.9%) | -3 | -1 | -3 | -14 | 1 |
| Poor Internal Organization | 7 (17.9%) | -1 | -11 | -5 | -1 | 2 |
| Minor Structure Issues | 7 (17.9%) | -4 | -1 | -5 | -9 | 0 |
| Low Cohesion | 6 (15.4%) | -1 | 1 | 6 | -4 | 6 |
| Duplicated Code | 3 (7.7%) | -8 | -3 | -23 | -39 | -21 |
| Too Many Temporary Variables | 2 (5.1%) | 0 | 0 | 1 | -2 | 0 |
| Unused Code | 1 (2.5%) | 0 | 0 | 0 | 0 | 0 |
| Poor Readability | 1 (2.5%) | 0 | 0 | 0 | 1 | 0 |
| Poor Method Name | 1 (2.5%) | 0 | 0 | 0 | 0 | 0 |
| Wrong Indentation | 1 (2.5%) | 0 | 0 | -1 | 18 | 0 |
| Poor Comments | 1 (2.5%) | 0 | 0 | 0 | 7 | 0 |
| Poor Parameter Layout | 1 (2.5%) | 0 | 0 | 0 | 2 | 0 |
| Long Method or Extract Method | 1 (2.5%) | -21 | -3 | -57 | -157 | -57 |
| Total (except defect category): | 39 (100%) | -38 | -18 | -87 | -198 | -69 |

In order to improve our capability of observation and data analysis, the refactoring effect over the metrics was normalized and its variation described through percentages. Percentage values were calculated thus: each metric variation value was divided by the metric value before refactoring. For instance, for a four-line reduction in a method with twenty lines we would have a -20% variation. After this normalization the data was analyzed using *boxplots*. Only the four most common categories were analyzed. For each *category* the potential outliers were kept in order to enhance the analysis considering that there were not many data points to be included.

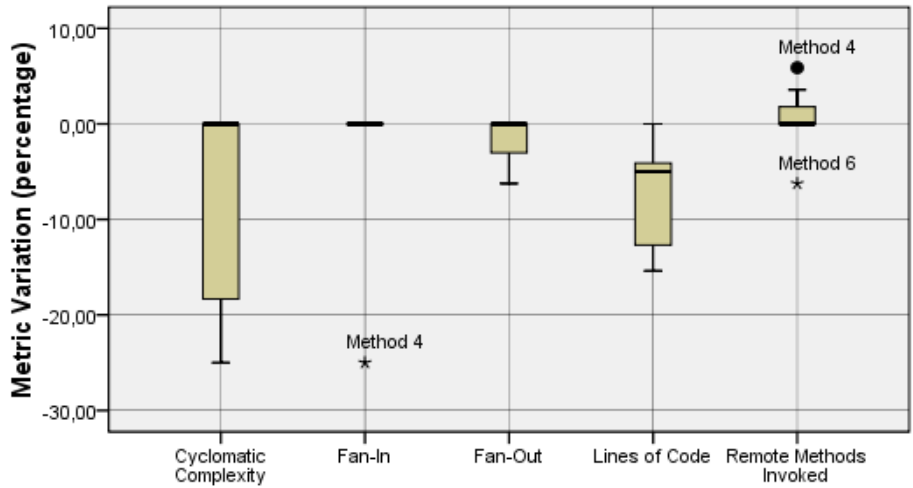

**Figure 6 - Metrics Variation Analysis for the Poor Algorithm Category**

The information shown in Figure 6 corresponds to the *poor algorithm* category. It is possible to observe the most affected metrics: cyclomatic complexity and lines of code. This behaviour can suggest that the perception of developers of what a poor algorithm represents is associated with an incorrect or unnecessary use of control statements, which is coherent with the general understanding of an algorithm that should be improved. Clearly evidencing this perception one of the developers used the following statement when describing the refactoring suggestion: 'Only the first 'if' is necessary, because if the search filters do not include the SSN, the 'else' of the second 'if' would appropriately create the list'.

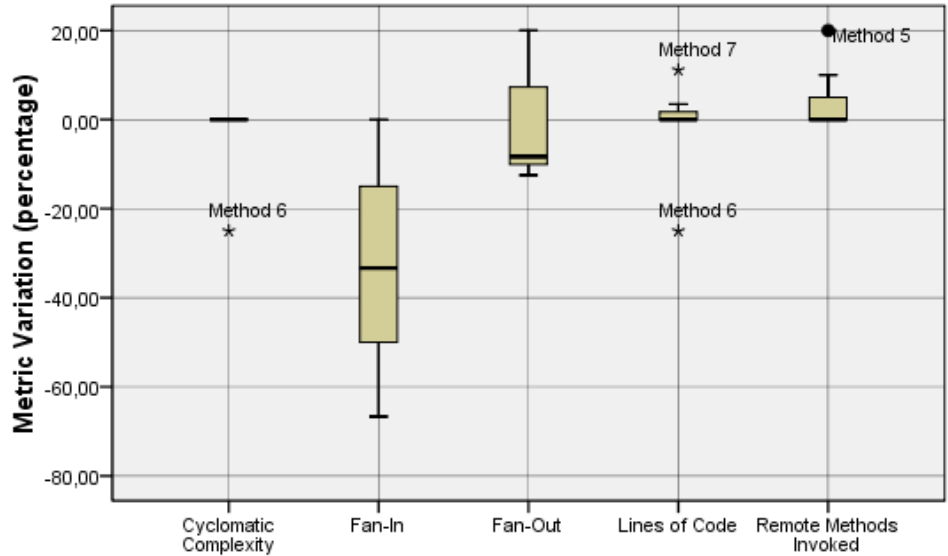
**Figure 7 - Metrics Variation Analysis for the Poor Internal Organization Category**

Figure 7 shows that the most affected metric in the *poor internal organization* category is fan-in. Indeed, more than half suggestions in this category involved recommending the correct variable scope use, as described by one developer: 'the reference variable 'searchForm' has its scope relevance only to this method and thus should be defined in it'. Other suggestions included recommending variables initialization in the constructor and avoiding the use of static text into the body source code (there is one special file in the project that should be used in case static texts exist).

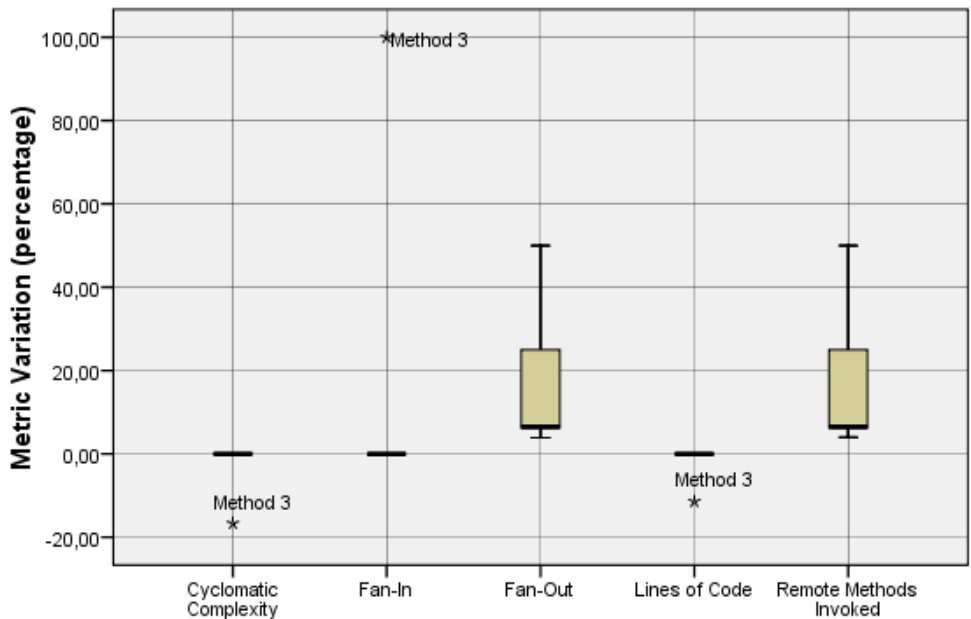
**Figure 8 - Metrics Variation Analysis for the Low Cohesion Category**

Most of the low cohesion refactoring suggestions were for violations of the MVC architectural style regarding the *low cohesion* category. The following developer description illustrates this scenario: 'the totalizing instalment method should not be in the 'form' class. It should be the responsibility of the 'InstallmentList' class.' The mentioned 'form' class is part of the View layer, which is basically responsible for maintaining the values filled in the forms. Figure 8 also shows an increase in the fan-out and remote methods invoked metrics. Nearly no variation in lines of code was registered.

The *minor structure issues* category included worries with source code organization such as assignment of unnecessary variables as pointed by one developer: 'an assignment to null is unnecessary in this case' or 'it is not appropriate to instantiate the 'Instalment' class without any state set and then consecutively invoke its 'sets' method. It is better to write a constructor that can receive all the parameters needed'. Due to its more generic nature, the *minor structure issues* category virtually had no significant metric variation pattern and presented a large number of outlier data points. For this reason, the *boxplot* analysis has been omitted.

### 4.4.2. Explaining the Refactoring Decisions

The metrics variation analysis presented in the previous section examined the effect over the source code metrics <u>after</u> source code refactoring. In this section, the source code structural characteristics <u>before</u> source code refactoring will be explored in an attempt to explain what metrics can have influenced the decision of the developers when suggesting a refactoring opportunity. This way, regression analysis was used to investigate how these characteristics could have affected the developers' decision. The categorical regression method was used in this analysis as it presents several advantages over traditional methods such as the ability to create combined regression models, where different independent variable types can co-exist and the dependent variable can be represented in any scale (nominal, ordinal, or ratio). Additionally, the categorical regression has better performance in a data set with few data points, several variables, and many different values for each variable. Due to these features, the categorical regression has been applied in this analysis. The accepted error level (*alpha*) is 5%.

For each category analyzed in the previous section, a regression model is created. The source code metrics was used as independent variables and the refactoring decisions for each method was used as dependent ones (using a nominal 'Yes', 'No' scale). The methods that did not receive any refactoring suggestion were considered as 'No' (10 methods in total) and those that received refactoring suggestions were considered as 'Yes' (the total depends on the category, although common quantities are 6 or 7 methods). Besides, the methods classified exclusively as defects were used as 'No' in the dependent variable to create the regression models. The refactoring focus did not regard the detection of defects. In case defects were not present, these methods could not have any impact on the refactoring suggestions.

Table 8 - Regression Model for the *poor algorithm* category

| Adjusted R Square: | 0.888 | p-value: | 0.018 |
|---|---|---|---|
| Model Predictors | | | |
| Predictors | Standardized Beta | p-value | Correlation Coefficient |
| Cyclomatic Complexity | -5.909 | 0.037 | -0.847 |
| Fan-In | -4.239 | 0.026 | -0.828 |
| Fan-Out | 4.324 | 0.078 | 0.904 |
| Lines of Code | 7.576 | 0.022 | 0.855 |
| Remote Methods Invoked | -4.154 | 0.058 | 0.910 |

Table 8 shows the regression model for the poor algorithm category, which seems able to explain 88.8% (Adjusted R Square) of the refactoring suggestions. According to the results, the most important predictor is Lines of Code, followed by Cyclomatic Complexity (we are considering the absolute value of standardized beta, as there some nominal scale variables; besides, the absolute value of the correlation coefficient defines the proportional variation of the dependent variable, not considering the influence of predictors' variables). Even when suggesting some Fan-in metric influence the model seems to be consistent with Figure 6, which shows that the same metrics were the most affected ones after refactoring.

Table 9 - Regression Model for the *Poor Internal Organization* Category

| Adjusted R Square: | 0.636 | p-value: | 0.047 |
|---|---|---|---|
| Model Predictors | | | |
| Predictors | Standardized Beta | p-value | Correlation Coefficient |
| Cyclomatic Complexity | -0.998 | 0.783 | -0.641 |
| Fan-In | 0.190 | 0.831 | 0.171 |
| Fan-Out | 3.160 | 0.091 | 0.603 |
| Lines of Code | 2.666 | 0.337 | 0.727 |
| Remote Methods Invoked | -5.115 | 0.012 | -0.752 |

The model shown in Table 9 involves 63.6% of the refactoring suggestions. The most important predictor is Remote Methods Invoked. Despite the model indicating the RMI metric as the most important regarding the developers suggestions, the most affected metric after refactoring (see Figure 7) was Fan-In. To investigate what could be influencing this difference the first step was to check whether the non-influence of the Fan-In predictor in the Table 9 model could indicate some similarity regarding the metrics values for those methods refactoring was applied to or not. One issue identified regarded the way one of the developers identified the problems in the methods. For instance, methods that were making use of large scope variables were not properly identified. This seems to represent the origin of the difference between the model in Table 9 and the behaviour seen in Figure 7. Even so, the model was able to capture the influence of the Remote Methods Invocation metric as also identified in Figure 7, although with greater intensity.

Table 10 - Regression Model for the *Low Cohesion* Category

| Adjusted R Square: | 0.676 | p-value: | 0.021 |
|---|---|---|---|
| Model Predictors | | | |
| Predictors | Standardized Beta | p-value | Correlation Coefficient |
| Cyclomatic Complexity | 0.581 | 0.916 | 0.421 |
| Fan-In | -0.213 | 0.941 | -0.191 |
| Fan-Out | -4.401 | 0.033 | -0.538 |
| Lines of Code | -1.417 | 0.605 | -0.729 |
| Remote Methods Invoked | 5.215 | 0.020 | 0.596 |

The most important metrics for developers' suggestions regarding the low cohesion category were Remote Methods Invoked and Fan-Out (Table 10). This model can answer 67.6% of the refactoring needs for this category and it is aligned with the behaviour seen in Figure 8.

It was not possible to create a statistically significant model for the minor structure issues category. In fact, this should be expected as this category produced the highest number of outliers, as previously discussed. In this case, this result is also consistent with the observed behaviour and no type of indication can be asserted.

To complement the data analysis, it is possible to observe through the metrics analysis before refactoring (regression models) that developer refactoring decisions are consistent with the effects in these metrics after refactoring. Therefore, as the source code metrics represent the structural characteristics of the source code it is possible to observe that the same characteristics influencing developers decisions were somehow directly affected after the accomplishing of the refactoring.

### 4.4.3. Perception on Code Quality and Learning – Qualitative Analysis

This section describes the qualitative analysis of the data collected from the developers throughout the semi-structured interviews (Section 4.2.4.2). Considering the goals of the study, it has been found that interview questions (iii), (iv), (vi), (vii) and (vii) (Table 5 and 6) should be the most significant ones as they relate directly to the results expected (questions Q.1 and Q.2).

The analysis will be supported by the use of Grounded Theory (Strauss and Corbin, 1990), with some adaptations from Baskerville e Pries-Heje (1999). According to Baskerville and Pries-Heje (1999), Action Research usually modifies the grounded theory element roles as it brings preset categories, possibly including a central category, which can be obtained from the research theme definition (Section 4.1.3) and from the initial literature review (Section 4.2.1). This way, the characteristic of grounded emergence of one theory is not genuinely reached as expected by the canonical grounded theory principles (Strauss and Corbin, 1990). Therefore, the so-called grounded action research (Baskerville and Pries-Heje, 1999) selects the grounded theory components according to the study objectives, focusing on the open and axial coding. Figure 9 shows the general canonical coding process overview. Its application in this study has been done until the Axial coding stage (represented by the Categories in Figure 9), based on interview transcriptions.

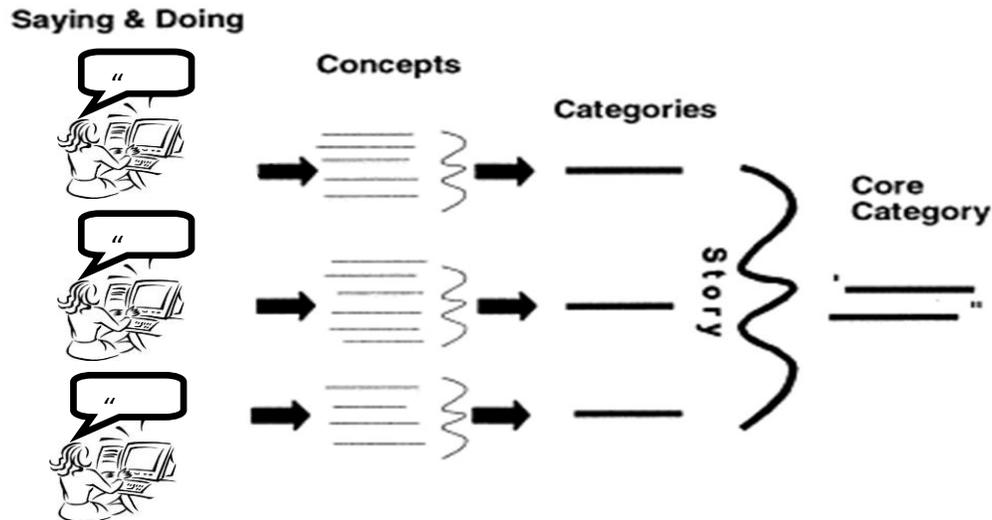

Figure 9 - Coding process overview – adapted from (Baskerville and Pries-Heje, 1999)

The main open coding objective is to reveal the essential concepts embedded in the data (in this case, represented by the statements made by the developers). So, the observations in the viewpoints of the participants are decomposed and organized into events or ideas by receiving meaningful labels (codes) (Baskerville and Pries-Heje, 1999). An additional open coding goal is to categorize the concepts, synthesizing them and supporting comprehension of the phenomena studied.

After performing the open coding process, we were able to get the results shown on Table 11. Each category was related to the research questions as defined in Section 4.2.2.2 as a way to represent their answers. The categories with IDs beginning with 'I' correspond to the ones extracted from the research theme. The association between categories and questions can be understood as the axial coding from the Grounded Theory. The relationships between the categories are related to the corresponding questions. At the end, the 'argument' column represents the link between the category and the concepts/codes discovered in adjacent data.

From the perspective of construct validity, it is important to mention that the research questions (section 4.2.2.2) allowed the definition of practical questions (section 4.2.2.3) and some of them associated to the semi-structured interviews (Table 5 and Table 6). This way, the trace from the research questions to the transcriptions can be observed, over which the Grounded Theory based analysis has been produced. This analysis approach intends to objectively demonstrate how the results were obtained.

Table 11 - Qualitative Analysis Categories (Grounded Theory)

| Id. | Question | Category | Argument |
|---|---|---|---|
| I1 | - | The source code refactoring process is analogous to source code inspection and easily applied under the perspective of the system dynamic execution according to the use cases models. | Developers when experienced using use cases artefacts can explore them to read OO code more efficiently (Dunsmore et al., 2003). |
| I2 | - | The refactoring process brings as benefit the learning regarding the defect patterns that can be documented through the results analysis. | Different studies suggest learning as one of the positive results of applying refactoring. |
| C2 | Q.1 | Tacit coding patterns are more present in the software architecture and source code structure (problem solution), representing essential knowledge for understanding the source code. | According to developers, knowledge regarding the architectural style and design patterns used in the software project allows to know *a priori* where to find the implementation of functionalities into a distributed OO code. |
| C3 | Q.1 | The more experienced developers are aware (and know) on the tacit patterns used in the software project and, therefore, can identify when they are being neglected. | The developers that applied refactoring in their own source code reinforced they were able to explicitly perceive (with the support of refactoring suggestions) a more uniform source code organization they did not consider before. |
| C4 | Q.2 | The learning concentrated on making explicit tacit reasons regarding the source code organization and internalization of coding patterns used in the software project. | A direct result from refactoring: the descriptions of refactoring suggestions presented the reasons as to why source code should be organized in a specific way and the participants could see defect patterns (in this case, through refactoring). |
| C5 | Q.2 | Knowledge regarding the architecture style and design patterns used in the project can be explained in the shape of directives that will guide developers in the use of a single development standard. | The participants understood that the directives would be the most adequate way to guide (objectively, specifically, and focused) novice developers in the software project. |
| C6 | Q.2 | All lessons learned shall be documented as categorized to facilitate their search and the reading of the directives. | The directives set can grow along the time. Categorization will help to keep the document focused in its proposal depending on the context (i.e.: directives related to naming conventions regarding domain classes) |

Despite the simplicity of Table 11 some comments are still necessary. Categories from Q.2 were able to support the learning stage in this study because they represented the basis to elaborate the documentation that is going to be used by the developers to learn on what previously was being considered tacit code patterns. Besides, the qualitative analysis and consequently all categories allowed to explore the different significance levels , which were built by the different actions conducted to serve as input to build local theories (Section 4.5.2.1) (Sjøberg *et al.*, 2008). The learning process and reflections on the results will be detailed in the next section.

## 4.5. Reflections and Learning

This section intends to explore the results of the study as project learning. This way, these results will be explored aiming at their organization and reflection on the knowledge acquired from the actions.

### 4.5.1. Learning

After the semi-structured interviews and data analysis, it was defined how the tacit knowledge concerned with the architectural style and coding patterns could be externalized. This externalization represents one of the main learning elements into the context of this study as it allowed the remote development team and future developers to familiarize themselves with the working patterns used in the software Project.

Knowledge externalization was accomplished by elaborating directives from the identified problems throughout refactoring. The directives goal is to inform on how to structure the source code and project, where each one of them deals with a particular issue. Besides the directives, the participants decided by a set of information that should be integrated to each directive aiming at allowing them to be used in practice. The defined set of information is:

- <u>Directive</u>: generic description (applicable to different contexts) for a rule regarding the source code organization and definition of a coding pattern.
- <u>Motivation/Example</u>: Explanations or examples on how the directive should be applied.
- <u>Arguments</u>: the reasons for the directive to be used.
- <u>Type</u>: directive category (related to the refactoring categorization but without applying a 'negative perspective', i.e.: poor readability was adjusted to be just readability.)
- <u>Impact level</u>: subjective evaluation regarding the influence on code quality if the directive is not applied. Three possible values can be used: 1 – low, 2 – medium or 3 – high.
- <u>Affected Classes</u>: types of classes where the directives can be more useful. Ex. Domain or utilities 'Controller' classes.

An initial group of directives was defined by the researchers and from that moment evolved with the contributions from all the developers involved with refactoring in the software project. Several improvements were suggested. For instance, developers suggested new types, examples, modifications in the descriptions of directives, amongst others. From the 45 detected refactoring needs (including refactoring opportunities + defects), 43 were used to define 23 directives. One directive can be related to one or more refactoring needs as some directives entail more than one refactoring need.

After the definition and evolution of the initial group, the directives were made available to the development team through a Wiki system already used in the software project, the TRAC system (Trac, 2003). The TRAC (by using a plug-in) allows the association of categories ('tags'), which can be indexed and searched. The Wiki represents an useful technology in this context as it allows the continuous improvement of the directives by the project team. As only a part of the source code was reviewed for refactoring it is possible that new directives will pop up when additional parts of the project are reviewed. For the sake of simplicity, only some examples are being presented here, including a description in Table 12 and a concrete directive instance in the TRAC system on Figure 10.

Table 12 - Directives Examples

| Directive | Motivation/Example | Arguments | Type | Impact Level | Affected Classes |
|---|---|---|---|---|---|
| Static Methods should be statically accessed, without instantiating the class containing the method. | Not correct:<br>Classe c = new NomeClasse();<br>c.metodoInvocado();<br><br>Correct:<br>NomeClasse.metodoInvocado(); | It not exist a need of creating a class instance to invoke a static method. Besides, memory space would be unnecessarily used. | Algorithmic Structure | 2 | All |
| Security verifications (facilities Access) shall verify if the grants of the identified user allow Access to one specific functionality. | Access Verifications were checking only the user role (and not the functionalities associated to the role). There are standard system roles capturing the permissions required for system actors, and specified in the use cases model. This model defines which actors can have access to the functionalities. The similarity between grants and actor role in project context allowed this misconception. | The User Management Module allows the maintenance of roles and grants by associating functionalities to the roles. This way, different roles can be created and have access to one same functionality. Therefore, access verification should be done considering the grants associated to the role and not just considering the roles. | Design Structuring | 3 | Controller/ Domain |
| Always possible (and needed) Boolean expressions shall be used as a method result. | return !(curso.isStatusEncerrado()); | In this case, the reduction of lines of code (in case of if/else constructors are being used) improves readability. | Readability | 2 | All |
| Always make use of the transaction management service when the persistence of the whole entity is needed or when the persistent class is responsible for persisting other classes. | GerenteTransacaoBD gtBD = new GerenteTransacaoBD(new ITransaction[ ] {proposta, projeto});<br>gtBD.iniciarTransacao();<br>try {<br>   proposta.save();<br>   projeto.save();<br>} catch (Exception e) {<br>   gtBD.recuperarTransacao();<br>   throw(e);<br>}<br>gtBD.finalizarTransacao(); | If there is no controlling and transaction restoring in the case of failure, inconsistent data can be incorrectly saved. | Design Structuring | 3 | Controller |

Figure 10 - Directive registering in the project's Wiki. The figure corresponds to the last example in Table 12: 1: directive; 2: motivation/example; 3: arguments.

Just after the agreement on the initial group of directives, all the developers recently integrated into the development team were asked to read the software project directives. One other expected benefit of this knowledge use is maintenance. It is expected that future maintenance will be accomplished by a totally different team and the directives represent useful knowledge for future interventions. The evaluation of this documentation usability is out of the scope of this study and represents an interesting work opportunity for future investigations.

### 4.5.2. Reflections

As previously mentioned, the main basis for the conducting of this study were the works by Mäntylä (2005) and Mäntylä and Lassenius (2006), which demonstrated how the study could be planned and organized to capture and analyze the quantitative (source code metrics) and qualitative (refactoring description suggestions) data, respectively. The intention is to compare or to comparatively interpret the results, even considering this study does not represent a concrete replication of previous Mäntylä's studies. A first comment regards the categorization used to identify the main refactoring factors. Considering this and previous studies, most of the categories are coinciding. Only 2 new categories were identified [considering the universe of 22 categories identified by Mäntylä and Lassenius (2006)]. This result reinforces the validity of the conceived categorization indicating these categories can appropriately represent the refactoring decisions. However, it also indicates the need for additional studies aiming at the verification of possible new categories. Besides, 11 categories from the original study were not identified in this work, which can indicate that the expertise of the researchers may have influenced the results.

The regression models prepared in both studies differ in some aspects and deserve to be detailed. The regression model based on the source code metrics described in Mäntylä (2005) was able to explain only 30% of the refactoring decisions in the original study. Using the same approach, it was not possible to create any statistically significant model in this study. However, for one group of participants in Mäntylä's study (2005) a list of pre-set, different types of design shortcomings concerned with encapsulation, data abstraction, modularity and hierarchy (called 'code smells' by Fowler et al., 1999) in object-oriented code (ex.: long method) was presented, where they were asked to look for the 'code smells' in the source code. Aggregating the models by 'code smells' allowed getting better performance and explaining 70% of the refactoring decisions for two or three 'code smells'. Using a similar approach in our study, after the categorization of the refactoring decisions, a regression model was created for some of the categories, with observable performance improvement and explaining from 63.6% until 88.8% in 3 from 4 categories.

Even so, there is some mismatch with Mäntylä's results (2005). In our study, source code metrics have been successfully used as source code refactoring decision predictors in one regression model. In Mäntylä (2005) the model was just statistically significant when considering the presence or absence of 'code smells' (not to decide on code refactoring). One possible reason for this difference can be associated with the experience of the developers in the project context, in our study. There was an expectation that practical knowledge could be useful to identify defects related to deviations in coding patterns and architectural styles. This way, there was consistency in the types of source code deviations identified, which possibly produced significantly better regression models for refactoring decisions.

There is some research exploring metrics to develop software technologies to detect refactoring opportunities in source code (Mens and Touwé, 2004); however, these technologies usually produce a lot of false-positive suggestions, requiring supplementary human evaluation in most of the cases (Parnin and Görg, 2008). Therefore, it is expected that the results presented in this work can support the creation of more effective tools. In fact, the use of regression models to reduce the number of false-positives on code metrics based analysis has demonstrated itself useful in near investigation fields, such as defects detection. Ruthruff *et al*. (2008) show how defects automatically reported by CASE tools (statically analyzing the source code by using source code metrics) can be filtered through regression models. These models can identify those reported issues having more chance to represent real defects. These regression models are usually based on manually filtered historical data.

This result can be used to motivate, for instance, the building of specific case tools to support the decision making regarding refactoring as these tools could allow self-calibration with real data from the software projects they are being used in. This data will include some historical subjective evaluation, according to the way the regression models have been created in our work. This way, the suggestion of Fowler *et al*. (1999) on keeping human evaluation in the detection of refactoring opportunities could be supported, even if indirectly.

Other work used to support our study (Dunsmore *et al*., 2003) defined a systematic procedure to read the different methods in the classes refactoring should be applied in. Considering both studies the developers reported that an important characteristic of this procedure is concerned with the possibility of thinking about the methods considering their execution contexts. An additional and important feature regards the fact the technique explores use cases models as the basis for its application avoiding the need for training as use case models represent a well know artefact commonly used in the software project.

However, despite their importance and influence, we could not relate all results to the aforementioned works of Dunsmore *et al*. (2003), Mäntylä (2005) and Mäntylä and Lassenius (2006). Two additional features have been explored in our study: (1) the use of refactoring as a means to explicit knowledge and document architectural styles and coding patterns for the software project; (2)

to compare the source code metrics based regression models with the observed metrics effects after refactoring.

The first feature represents the learning dimension of our study. Therefore, we believe it can represent important evidence regarding the possibility of inspections (reading), and can also support the externalization and documentation of tacit knowledge on architectural styles and coding standards.  The refactoring process allowed the development team to think about its previous experience and, while describing the refactoring needs, to state the reasons why refactoring was considered necessary. Besides this indication, different quantitative and qualitative analysis moments were able to demonstrate the concrete existence of tacit knowledge that could impact the quality of the software being produced. For instance, the quantitative analysis after considering the existence of tacit knowledge has highlighted the same metrics that could be influenced by the developers' decisions regarding refactoring. It can indicate developers were conscious and consistent in their decisions based on their (tacit) knowledge. It can be seen by observing the impact caused in the same set of source code metrics.  Regarding the qualitative analysis, apart from allowing the identification on how the lessons learned should be organized and formatted, it revealed the way tacit knowledge can emerge and be moulded into the software architecture and source code structure as a collaborative and cooperative development work result.

*4.5.2.1. Theory Construction*

Aiming to systematically organize and structure the series of events and knowledge concerned with this study, we decided to ambitiously represent them as a theory. The use of theories in Software Engineering is still uncommon, but due to the richness of the observations from action interpretations the building of such theory could facilitate the communication of ideas and generated knowledge (Sjøberg *et al*., 2008). Apart from that, the preparation of the theories represents an interesting strategy to facilitate its future aggregation through secondary studies (Charters *et al*., 2009), as it can clearly explain the applicability and limitations of the investigated software technology. According to Sjøberg *et al*. (2008) there are three different levels of sophistication or complexity regarding the theories:

- <u>Level 1</u>: small stable and concrete relationships based on direct observation;

- Level 2: mid-range theories with some degree of abstraction but still heavily connected to the observations;
- Level 3: general theories looking to explain phenomena in the context of Software Engineering.

These levels establish markers in the generation of theories but can also represent complete theories by themselves. According to Sjøberg *et al*. (2008), the development of theories in Software Engineering should be initially focused on levels one and two. The theory elaborated in this work can be considered a Level One theory.

Sjøberg *et al*. (2008) suggest the description of theories should be split into four parts: constructs (the basic elements), propositions (how the constructs relate), explanations (why the proposition was specified) and scope (what the universe of discourse in which the theory can be applicable is). This way, the construction of theories should involve 5 steps: (1) definition of constructs; (2) definition of propositions; (3) providing explanations to justify the theory; (4) defining the scope of the theory, and; (5) testing the theory through experimental studies. Sjøberg *et al*. also state that the use of a grounded theory can facilitate the identification of the main concepts (constructs) and their relationships (propositions and explanations). As a consequence, the constructs that will be presented have been based mainly on the grounded theory analysis conducted in our study.

Figure 11 shows the graphical theory schema (Table 13 presents the theory elements in detail). The notational semantics has been defined by Sjøberg *et al*. (2008) and is explained below. A construct is represented as a class or class attribute. A class is represented by a box with its name written at the top, such as, for instance, 'Distributed Project'. A class can have a subclass (using the same generalization notation as in UML) or a component class (drawn as a box inside another box such as, for instance, 'Source Code'). Usually, if the construct represents a particular variable value, then the construct is modelled as a subclass or component class. However, if the focus concerns the values variations, then the construct is a variable modelled as a class attribute, such as 'Effort'. An attribute is described as text in the class box bottom (below the horizontal line). All the constructs are underlined in Figure 11.

A relationship is modelled as an Arrow; an Arrow from A to B means that A affects B, where A is a class or an attribute and B is an attribute. Considering a relationship, B can also be a relationship in itself, also represented by an Arrow. In this case, A is called a moderator, as in the case of 'Experience' in Figure 11. It means A affects the direction and/or intensity of the B relationship effect, so the moderators can be defined as propositions (see proposition P7 in Table 13).

The inheritance bases for all the classes are called archetypes (actor, technology, activity, and software system). According to Sjøberg *et al*. (2008), the typical scenario in Software Engineering can be represented as an actor that applies a technology to support some activities in one software system (planned or existing). Some examples of an activity archetype class are creating, modifying, and analyzing (details in Sjøberg *et al*. (2008)).

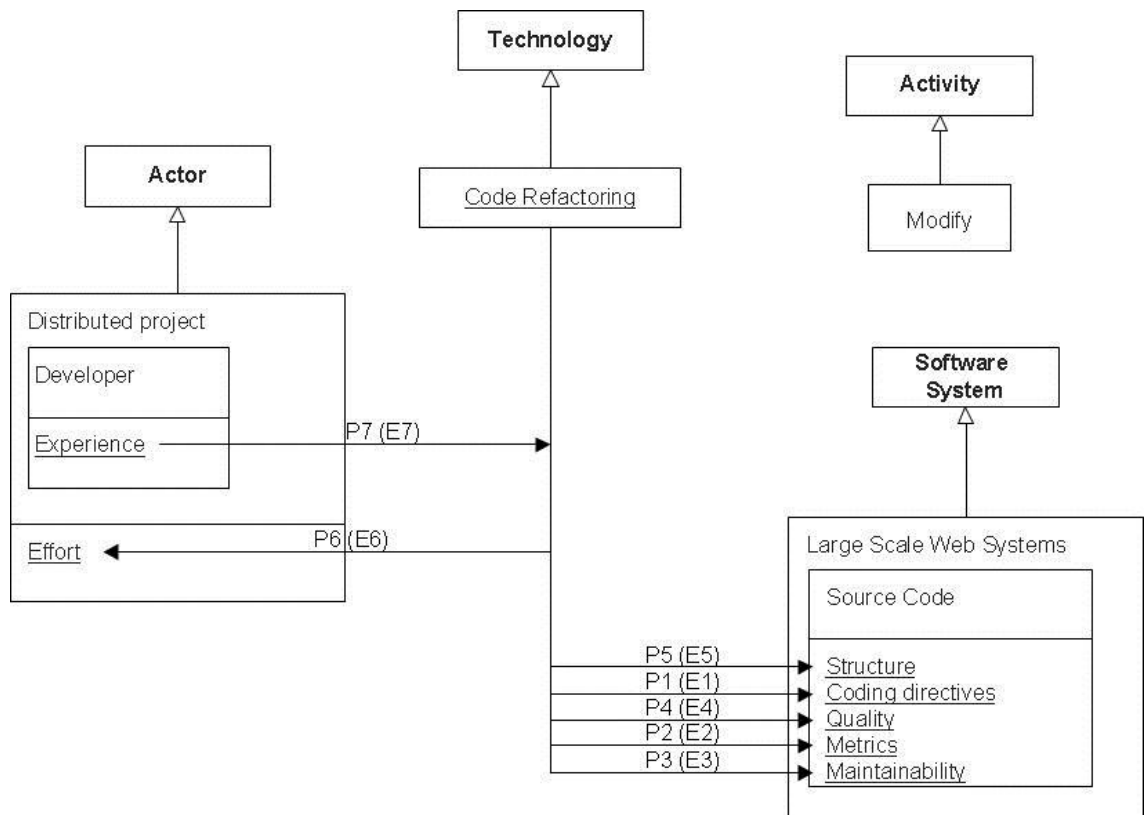

**Figure 11 – Theory diagram on the use of source code refactoring to explicit coding conventions in a medium-to-large scale Web software project**

**Table 13 – Description of theory elements**

| Constructs |
|---|
| C1 *Code Refactoring (*based on the reading technique of use cases by Dunsmore *et al*. (2003) and registering the refactoring suggestions according to the work of Mäntylä and Lassenius (2006)) |
| C2 *Source Code Structure (*structural properties perceptible in the source code, ex.: readability, algorithm structure*)* |
| C3 *Source Code Quality (*number of defects per lines of code*)* |
| C4 *Source Code Metrics (*lines of code, number of parameters, cyclomatic complexity, number of invoked remote methods, 'fan-in', 'fan-out'*)* |

| | | |
|---|---|---|
| C5 | *Coding Directives (*coding Standards and architectural styles*)* | |
| C6 | *Maintenance Facility (*regarding the effort needed to maintain the source code*)* | |
| C7 | *Developer Experience (*developer involvement time in the project where the source code will be reviewed for refactoring*)* | |
| C8 | *Effort (*total man-hours/hours allocated to project activities*)* | |

**Propositions**

| | |
|---|---|
| P1 | Source code refactoring positively influences coding directives |
| P2 | Source code refactoring positively influences source code metrics (value reduction) |
| P3 | Source code refactoring positively influences source code maintenance (facilitates) |
| P4 | Source code refactoring positively influences source code quality |
| P5 | Source code refactoring positively influences source code structure |
| P6 | Source code refactoring negatively influences software project activities effort (increase) |
| P7 | The effects of code refactoring are reduced when the developer does not have previous experience with the software project context. |

**Explanations**

- E1 The coding directive can be constructed or evolve.
  - The descriptions of refactoring suggestions explain the way the source code was structured
  - The directives are directly derived from the refactoring suggestions
- E2 Source code metrics values reduce
  - The measures usually display a reduction after refactoring, mainly because the main goal of refactoring focus on improving source code features such as readability and performance.
- E3 Source code maintenance is facilitated
  - The coding directives allow *a priori* understanding on how the source code is organized and, therefore, facilitate its understanding when some maintenance activity is needed.
  - The source code structure becomes more homogeneous which can ease future maintenance.
- E4 Source code quality improves
  - Developers identify source code defects (even when it does not represent main refactoring focus, source code inspection is being applied).
- E5 The source code structure improves
  - The source code structure becomes more homogeneous throughout the entire software project, according to the previous knowledge of the developers.
  - The size and complexity of the source code is reduced.
- E6 The project effort increases
  - Need to allocate resources (people – man/hour) to accomplish the activities related to source code refactoring.
- E7 The developers experience in the project allows
  - Capturing and reasoning on the coding standards used in the project
  - Identifying deviations in coding standards during refactoring
  - Identifying defects related to requirements and domain knowledge
  - Executing refactoring activities in less time.

**Scope**

This theory is supposed to be applicable to software projects with distributed teams (same native language and time zone) creating and modifying medium/large scale Web Information Systems on Java based development platforms and using incremental development supported by VV&T activities.

## 5. Applying Action Research to Software Engineering

If properly planned and executed, the Action Research methodology with its dual objective of improving organizational problems and generating scientific knowledge leads to a 'win-win' scenario for both professionals (organization) and researchers.

The primary goal of Action Research is represented by its self-changing process that is not only observed by the researchers but also influenced by them. This makes Action Research especially appropriate to investigate not homogenous through time phenomena (Checkland and Holwell, 1998) that are not reproducible. Usually, these phenomena are social events such as most of the Software Engineering activities. Hence, in order for Action Research to be well conducted and provide relevant results, it imposes the researcher an additional set of knowledge items and abilities (Baskerville, 1999). First, it is necessary that the researcher holds a deep knowledge of the organization's processes and organizational culture. Secondly, the researcher should be able to interpret and understand the field under observation; plan and conduct the interventions; collect, analyze, and construct the intervention data; formulate concepts and theories, and prepare theoretical explanations; and establish collaboration with people and the organization, including the dealing with ethical issues.

Most of this knowledge and abilities required by Action Research are constantly demanded during the execution of the research because the proposed solution that is being implemented depends directly on the decisions of the researcher. These decisions are crucial not only to allow generating genuine scientific knowledge but also to conciliate this with the organization's business needs. They are even more decisive if one considers that in Action Research the intervention occurs in a real environment where *in vivo* study results can directly affect the organization. Add to this the fact that, contrary to controlled studies, the observed object is not promptly available but is elaborated and manipulated during the research with the participants. This way, even though careful planning has to be considered in Action Research studies its reach is limited by the improvisations made by the researchers in their daily activities.

However, although heavily challenged by different demands faced in its activities, a researcher initiating with Action Research is usually only supported by generic descriptions and diagrams (as shown in Section 2, which are commonly found in the technical literature) that can lead to a vision far from what Action Research actually represents (McKay and Marshall, 2007). Indeed, several researchers argue that Action Research is more closed to a notion of strategy rather than a methodology or technique (Heatwole *et al.*, 1976). Some researchers even question whether Action Research could not be considered a research paradigm (Lau, 1999). In consequence, Action Research is usually defined and presented as general recommendations for, as an example, simultaneously performing action and research, collaborative research, and learning by reflection. These different perspectives for Action Research leave open important issues in terms of knowledge and abilities necessary to its practice in specifics disciplines such as Software Engineering.

Thus, it is not a coincidence that several disciplines have already addressed this theme, including Nursing (Holter and Schwartz-Barcott, 1993, Meyer, 1993), Political Research (Heatwole *et al.*, 1976), Management (Ottosson, 2003), Operations and Production Management (Westbrook, 1995), Logistics (Näslund, 2002), Marketing (Ballantyne, 2004, Kates and Robertson, 2004) and Communications and Media Research (Hearn and Foth, 2004). Generally, these studies can be categorized into two types. The ones that discuss the potential benefits Action Research can introduce to their disciplines (Heatwole *et al.*, 1976, Holter and Schwartz-Barcott, 1993, Näslund, 2002, Kates and Robertson, 2004, Hearn and Foth, 2004). This is what we have aimed at in Sections 2 and 3. And those that propose Action Research adaptations and guidelines according to the particularities of their areas (Meyer, 1993, Ottosson, 2003, Westbrook, 1995, Ballantyne, 2004). This is what we intend to focus on in this section.

In the areas above that have employed Action Research we can clearly see particular focuses on the use of Action Research. For instance, Meyer (1993) has a special interest in the ethical issues concerned with the consent of the participants in a context marked by a large number of different specialists and intense turnover of the professionals as in the case of several nursing wards; Ballantyne (2004) points out the importance of a collaborative spirit establishment via Action Research as a foundation for the formulation and dissemination of marketing strategies amongst the organization staff; and Hearn and Foth (2004) highlight the ability of Action Research in revealing tacit knowledge embedded in an organization , which wants to expose its identity in new communication media forms.

Therefore, based on the way Action Research has been applied to these areas, we gathered a set of practical issues that deserve to be discussed in order to guide and foster Action Research studies in Software Engineering. All of these issues will be addressed in the following sections:

1) What are the essential abilities software engineers should have to apply Action Research?
2) What are the relevant Action Research aspects in the research process management?
3) How can the organizational culture influence reflection and collaboration?
4) What are the possible ethical conflicts generated in conducting an Action Research study?
5) How can new (scientific) knowledge be generated?
6) What are the benefits of Action Research for the construction of theories?
7) How can the learning be extracted from the interventions?
8) What are the main formats Action Research can have in Software Engineering?

## 5.1. Practices and Discussion

### 5.1.1. Learning and Difficulties in Using the Action Research Methodology

Starting an Action Research study in Software Engineering is no trivial task. One first hurdle to be overcome by the researcher is the lack of guidelines on how to use Action Research in Software Engineering. The absence of technical papers on this theme can lead the researcher to an uncomfortable situation where one should not only understand the Action Research methodology but also find a way to apply it in the field.

After understanding the main benefits and limitations of the methodology in Action Research the researcher should then evaluate whether it is really appropriate to one's research purposes. At this point, it is important to be certain that the principles of Action Research can be met: changing through action and learning by reflection. Still, in certain cases, it can be difficult to perceive the difference between an Action Research study and a Case Study. This difficulty is even more noticeable when there is more emphasis on observation than on action at certain moments. Nevertheless, if there is an intention to change the organizational culture and a concern with the wide consciousness of its employees on the situation faced and the solutions adopted, then Action Research can be considered as a feasible option for an investigation methodology.

However, in many cases, the assessing of the suitability of Action Research for research purposes cannot be done before a precise situation diagnosis is made or even before the planning stage. This occurs because only in the diagnosis stage the researcher has the chance to examine in detail the problem being addressed and take an informed decision on whether Action Research fits the problem investigation. Similarly, only during the planning stage can the researcher create solutions and assess if Action Research is appropriate for its conclusion. For instance, over the refactoring study meetings described in the previous section where one of the authors played both professional and researcher roles, the experienced developers in the project saw quality issues in the source code produced by the remote team, which had recently been integrated in the project. In moments like this, the researcher should have a refined sensibility to perceive a research opportunity. From the scientific point-of-view, this sensibility is associated to the awareness of the state-of-the-art of the theme under investigation as well as an understanding of the scientific methodology, especially that concerning the reflexive and critical thinking (e.g., formulation of hypotheses and their potential answers to guide how the intervention will be driven). From the point-of-view of the practitioner, it is associated with a refined technical knowledge on the problem being addressed to allow the researcher to adopt a pragmatic stance based on one's experience. Thus, considering these preconditions to start an Action Research study and the lack of guidelines for its application in Software Engineering, the low number of Action Research studies can be justified, as described in Section 3; especially when considering that software engineers usually do not have the habit of being reflexive of their actions (Rus and Lindvall, 2002).

Consequently, it is up to the diagnosis and planning stages to enable researchers to develop their initial feeling on the real need to scientifically investigate a problem. These stages enable the identification of primary causes and circumstances faced by an organization and to propose the design of a possible solution. In the refactoring study the diagnosis allowed hypothesizing the existence of tacit knowledge based on the analysis on how the teams were structured and its implications to the remote communication amongst its members that possibly led to the problems that are detected (poor source code quality). However, only during the planning stage (more specifically in the technical literature survey) it was possible to check that the theme had not been much explored yet. Indeed, no study has specifically addressed the circumstances faced, but the researcher (by using technical knowledge and based on the studies reasonably related to the problem) was able to choose the source code refactoring as a feasible way to solve the problem. It is important to note that until the planning the research theme had not been completely defined (i.e., using source code refactoring as a mean to externalize tacit knowledge associated with the architectural style and coding conventions). Consequently, the diagnosis stage needed to be revised, regarding especially the definition of the research theme. This characterizes an important Action Research process property regarding the possibility of iteration between stages.

Another important aspect that deserves attention during planning is data collection. The previous definition of what data should be collected shields the researcher from unexpected events allowing one to focus on the monitoring of actions and the participating in interventions. Obviously, unplanned events will happen but the Action Research methodology can accommodate this kind of deviation. Nevertheless, it still necessary to register these events to allow the researcher to justify them. It gives greater rigour to the relevance of both the research and the results. As an example of this kind of deviation in the refactoring study, we initially planned that each developer would review all the source code related to all listed use cases but later decided that the best strategy would be to split the work between the two developers in order to save project resources. This separation was only possible because the categorization of refactoring needs allowed data aggregation regardless of the reviewer. Hence, it is important to note that this decision was taken only after ascertaining it would not impact data analysis.

Still concerned with data collection, it is also necessary to observe how to proceed with the collection. Automated forms (e.g., voice recording) provide the researcher with additional time to conduct the interventions. Even if at a first glance this appears to be a trivial issue, in the refactoring study this was not initially planned and caused the interviews with the developers to take longer than necessary and planned. In studies with a large number of participants this issue can be even more critical.

As a final remark regarding learning it was interesting to observe how Action Research can accommodate the use of different types of data (quantitative and qualitative) and data analysis techniques. This reinforces the methodology flexibility in achieving its main goals: diagnosed problem solution, organizational learning, and scientific knowledge generation.

**5.1.2. Managing Research Interventions: Collaboration + Tacit Knowledge + Reflection = New Knowledge Opportunity**

Due to its characteristics Action Research is dependent on the collaborative participation of the involved people to allow the evidencing of tacit knowledge on the know-how of the performed interventions. Thus, with the analysis of/reflection on these actions one can identify relevant phenomena that show the interrelationships (seldom of cause-effect) between the solution and problem.

Different factors can affect the intensity of collaboration in an Action Research study. Being a research with, and for, the professionals in the investigated organization, the management of the research actions requires special attention with the selection of the participants, who should be engaged to take part in the research, to allow them to genuinely collaborate to achieve the global goals. It would also be convenient if a good professional relationship existed amongst them, to allow creating a less constrained environment for the emergence of new ideas, facilitating spontaneous communication and the dissemination of knowledge throughout the organization. Evidently, these conditions cannot always be met. To overcome the risks, it is important for the researcher to have diplomatic abilities in getting support from upper management to implement the intended actions. According to Ballantyne (2004), it is also important that the researcher shows commitment towards the set goals, acting on the same condition level of other participants in an attempt to create an atmosphere of trust and giving confidence to them in challenging the established organizational culture with their actions. In other words, the behaviour of the researcher is important to indicate that one is on the same 'side' of the participants and that one should not be intimidated by upper management. In the refactoring study this was naturally achieved as the researchers already were organization employees.

All the participants enter an Action Research study as apprentices even if they are specialists in their working areas. Nevertheless, a great variety of knowledge and skills are crucial and can include not only their direct experiences but also indirect ones (e.g., industry consensus or advisory companies). This type of knowledge is frequently named tacit knowledge and represents the know-how professionals bring in to perform their tasks. Rus and Lindvall (2002) argue that the software development process is a design process where all involved have a great number of decisions to make and, in order to develop complex software systems, they need to communicate amongst themselves so that the individual (tacit) knowledge can be disseminated and used in the project, and by the organization as a whole. This way, communication plays an important role in the research as it is the conducting medium that enables collaboration amongst the participants.

During the studies we conducted, all of these elements, i.e., collaboration, tacit knowledge, and dialogue, were widely explored. The refactoring study, for instance, started with hypothesizing the existence of tacit knowledge on coding conventions (i.e., practical knowledge) that could be made explicit through the process of source code refactoring. Therefore, the first thing done in that study was trying to make the software developers indirectly reflect on their practices while they identified and registered the coding conventions established by them in an unplanned way. In order to facilitate this process we tried to avoid an unfriendly disposition between the local and remote teams by informing the remote team that quality assurance was always a major concern in the project and the responsibility for the problem faced was not exclusively theirs but instead of both teams, in learning how to work remotely. The researcher should always be aware of these types of organizational and interpersonal issues. Due to this configuration and the overall study organization the problem was naturally faced as a normal project activity, having the dialogue between the developers flow spontaneously. The dialogue background consisted of the recorded refactoring opportunities. From these descriptions, the teams could interact exploring their own points-of-view and, supported by the researcher, decided how they could demonstrate tacit knowledge, and what coding conventions were more relevant. All these decisions were also documented and later disseminated to the rest of the team. To summarize this process, and putting the importance of collaboration and dialogue in a few words, Schein (1994) argues that dialogue is the form of 'thinking together to construct shared meanings'.

From all the manipulation of tacit knowledge used by the developers and the disclosure of their behaviours during the execution of the activities, the reflection (characterized by learning in Action Research studies) represents an immediate outcome from the interventions. Clearly, in many cases additional effort will be needed to format the outcome before it can be disseminated to the organization. It was the case in the refactoring study when we decided to use the TRAC system (Wiki-based system). However, the most important is that the organization becomes aware of how the circumstances were faced and the problem solved in a way that it can be avoided in the future. Further, it is up to the researcher to inform the participants on the results from the academic point-of-view and how it relates to their learning. These academic results aim at demonstrating that interventions to achieve effects represent new scientific knowledge, by comparing them with other studies published in the technical literature, always exploring collaboration and the rich meaning layers of tacit knowledge that Action Research can exploit.

Based on the research process under discussion it is possible to realize how the collaboration, tacit knowledge and reflection are essential elements for the practice of Action Research, in enabling the building of new knowledge. Indeed, Rus and Lindvall (2002) note that knowledge in an organization can be created/rise from learning, problem resolution, innovation, or creativity. Thus, Action Research can be thought of as a means to allow the process of knowledge creation by working not only in the organizational context but also trying to manage it under a scientific perspective. Theory building can be useful to the latter objective as we intended to demonstrate with the refactoring study.

### 5.1.3. Ethical Issues

Being a collaborative research in a real context, the ethical issues involved in an Action Research study become even more important. From the researcher's point-of-view, it can be difficult to pursue their concurrent scientific goals, especially considering the requirements to publish in journals and conferences. This way, this can be an ethical issue that researchers should not put the problem solution in an unfavourable position in relation to their academic interests as the initial agreement between the parties (researchers and organization) should always aim at addressing a real problem. Researchers should try as much as possible to fit the research tasks as normal project activities and what is not possible to be addressed should be registered to later analysis. This reinforces our prior discussion on the importance of using automated mechanisms to collect data that can allow researchers concentrate on the solution of the problem. As a consequence, even though the Action Research study is conducted in 'real time', most of the analysis is usually retrospectively executed. In our studies we used several automatic or even 'implicit' data collection instruments such as voice recording, source code metrics and inspection software tools (the use case inspection study was not described in this book chapter).

A second issue regarding ethics concerns the participants' agreement in taking part of the research. As it is conducted in the organization they work for and as it deals with a real problem, those involved feel compelled to participate as several research activities overlap their professional responsibilities. To alleviate this feeling it should also be offered to these professionals that their individual data will not be published even though they will be participating one way or another. Hence, the freedom in consenting cannot be always achieved and once the project has been initiated it becomes even more difficult to the professionals to feel comfortable in making their decisions as they are already involved in the process. Therefore, it is important to reinforce that the researcher should try to act on the same condition level of other participants so that they can get the impression they are being supported in their activities.

Apart from these issues, the researcher should not put common ethical issues away from any research effort, such as honesty, the privacy of the participants, amongst others.

## 5.2. Recommendations

### 5.2.1. Engineering Software and Building Theories

As discussed in the beginning of this chapter, even today the adoption of *ad-hoc* solutions throughout software development is still common because of the pressure produced by the time to market or lack of knowledge. Unfortunately, this goes against the Software Engineering definition given by Barry Boehm in 1976 (Boehm, 1976): 'the practical application of scientific knowledge in the design and construction of computer programs and the associated documentation required to develop, operate, and maintain them'. Yet, Action Research can bring this initial expectancy on software engineering into reality as it is designed to use and apply scientific knowledge in real contexts even where there is pressure for quick results, due to its dual research and action goals.

Action Research accepts that scientific results, even in their preliminary stages, can be used in practice for an immediate feedback on their value. Additionally, the Action Research methodology allows an *ad-hoc* solution to be collaboratively developed, even when no scientific results are directly related to the problem faced. But, contrary to the common practices in Software Engineering, special attention is given to reflecting on, and learning from, the results achieved. Figure 12 illustrates this process.

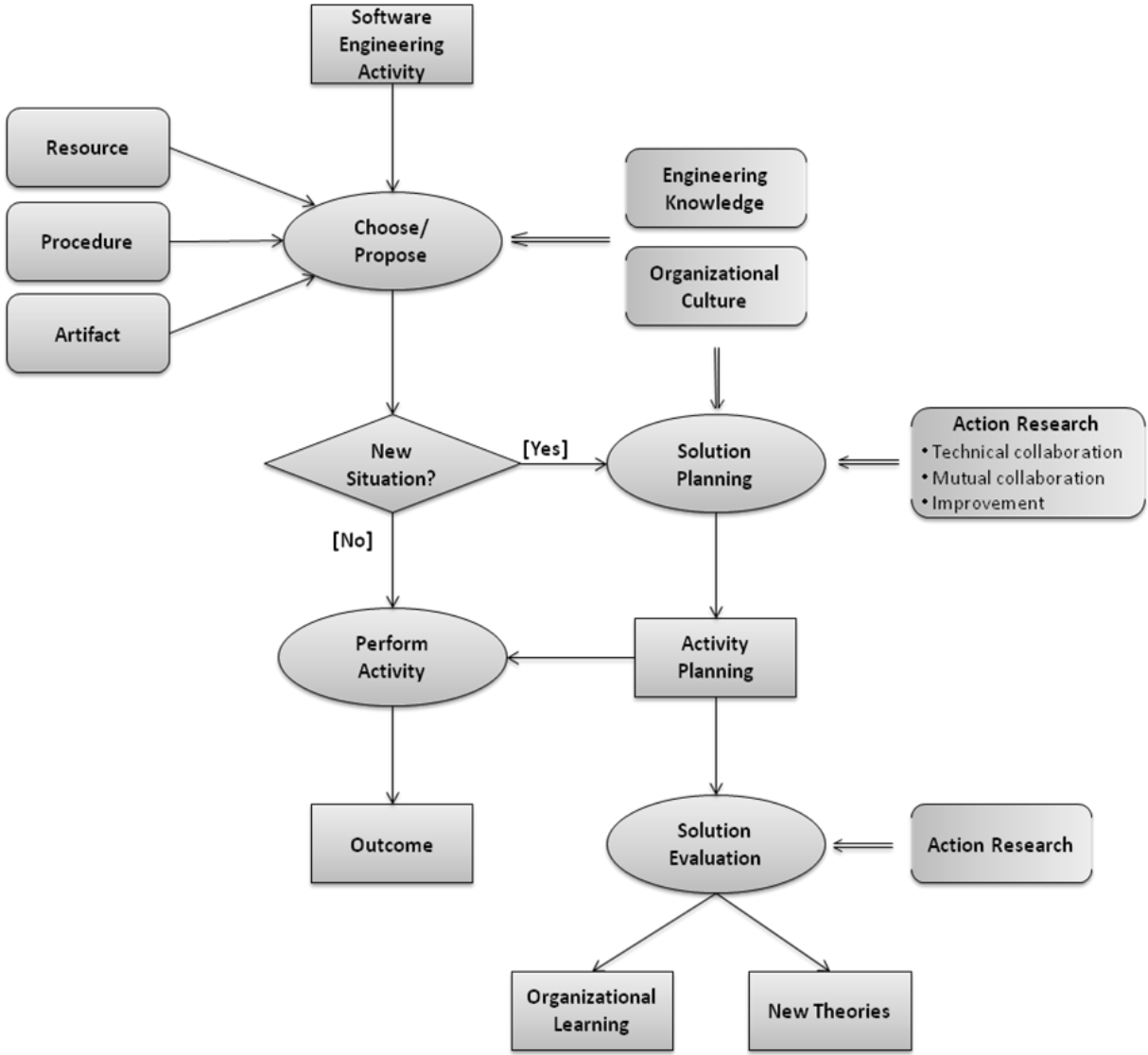

**Figure 12 – Construction of new theories through Action Research in Software Engineering**

According to Falbo *et al.* (1998), a software process can be characterized by the activities planned to be executed and the resources, procedures, and artefacts whose use is planned. In each activity, the resources, procedures, and artefact should be defined to produce the desired product, supported by an engineering reasoning and influenced by the organizational culture (Table 14).

**Table 14 – Key terms in the practice dimensions of Software Engineering – adapted from Higgs *et al.* (2001)**

| |
|---|
| **Evidence based practice** can be thought as justifying the engineering decisions on the best available evidence. |
| **Engineering reasoning** is the decision-making process associated to the application of different techniques and scientific principles to define a software technology that is properly designed and economically viable so that it allows its final conception. |
| **Engineering decision making** is a sequence of judgments made by engineers when interacting with the environment considering a set of restrictions (resources, material resistance, schedule, amongst others) |
| **Engineering judgment** consists of balancing available evidence with knowledge on the context and domain. |
| **Propositional knowledge** is derived from research and theory. |
| **Expertise knowledge** has its origins associated with a rigorous assessment and processing of the professional experience. |
| **Personal knowledge** results from the personal life experience. |

However, at some point in the software process the knowledge on software engineering can be insufficient in a given situation (e.g., it may not be possible to know what procedure to adopt in one specific software design activity). In this case, the Action Research methodology can be used to address the problem in a pragmatic way without excluding scientific rigour. As shown in Figure 12 the first step is to plan how the activity will be conducted by searching, as an example, for a procedure that is provided by the scientific community, or choosing to develop it within the organization itself. Then, the activity is regularly performed while, at the same time, its execution will be evaluated so to subsidize organizational learning and the building of new theories. The three Action Research formats we identified in our technical literature survey as shown in Section 3 were named technical collaboration, mutual collaboration, and improvement, based on the nomenclature provided by Holter and Schwartz-Barcott (1993). The first one focuses on testing a technology in a real context; in the second one researcher and participants jointly identify the problems, their causes, and possible interventions; and the last one aims at facilitating Software Engineering activities. We can rate our refactoring study under the second format. But most important is to notice how the role of the researcher changes between the formats, acting as observer in the first, participant in the second, and facilitator in the third.

Nevertheless, to identify when a new problem is being faced or, in other words, detecting a genuine research opportunity may not be an easy task. Given the constant flow of events and context variables found in a real environment, these opportunities can be hidden or can be incorrectly identified. As a basic rule, the research theme at hand should not have a trivial solution or, to put it in another way, it should be something that is not available in the industry as a common solution to the problem. As an example, in the refactoring study, the documentation on the tacit coding conventions and architectural styles used in the project could not be automatically performed by a tool, because knowledge was tacit. Moreover, systematic procedures to produce this documentation were not found either.

The organizational culture can also be an inhibiting factor in identifying new research opportunities or even become a barrier to resisting change. Normally, it is related to the beliefs of individuals, as well as the actions and practices of the organization (

Figure 13) (Kates and Robertson, 2004). In one of the studies we conducted regarding the inspection of the description for use cases (Santos and Travassos, 2010), several discrepancies pointed out by the checklist used were initially rejected because the participants argued that the use cases were being specified in that way since the beginning of the project and thus they were not seeing any problem in what had been indicated as discrepancies. Hence, more than hampering the identification of new research opportunities, the organizational culture can be the source of the problem that is faced as it becomes the basis for incorrect affirmations.

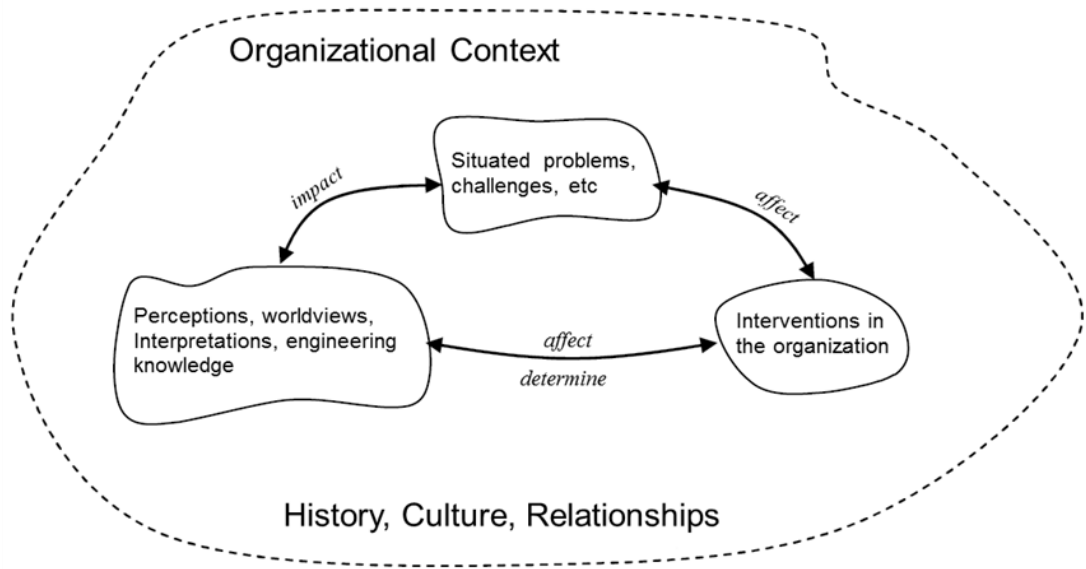

Figure 13 – Organizational context in Software Engineering – adapted from McKay and Marshall (2007)

The researchers should deal with this context by using tacit knowledge and exploring relationships to their advantage, to tackle the challenges and promote new actions and interventions in the organization without putting aside their history, whilst attempting not to cause disagreement. The role of the researchers is thus to conduct the interventions as naturally as possible, trying not to cause major disruptions. If they can fulfil this role they will be able to manipulate the worldview of the participants towards the proposed interventions goals and reproduce the phenomena identically to the usual daily practice of the organization, contributing to an improved relevance in the results, for the organization and for the research. In order to make this happen as naturally as possible, researchers should also consider the software engineering practice dimensions previously mentioned. To support the research in this issue we suggest the mapping between the dimensions and Action Research provided in Table 15.

Table 15 – Mapping Action Research and Software Engineering practice dimensions

| Action Research | Software Engineering practice key terms |
| --- | --- |
| Diagnosis | Personal knowledge and expertise. |
| Planning | Evidence-based practice, engineering reasoning, propositional and personal knowledge. |
| Intervention | Engineering reasoning, personal, expertise and propositional knowledge. |
| Evaluation | Propositional knowledge. |
| Reflection | Personal, expertise and propositional knowledge. |

Figure 14 illustrates how this research process fits into an organizational context. As previously discussed, after data collection, the researcher can organize it for further analysis and then conduct the analyses using a technique that can produce knowledge, serving as input for organizational learning and to produce scientific knowledge, possibly in the shape of a theory. The dash lines in Figure 14 show the main source of constructs, propositions, and explanations that form a theory. During the manipulation and organization of the data collected, the researcher can recognize the central concepts and objects used and affected by the research that will become the immediate candidates for constructors. Since most part of the collected data in an Action Research study is qualitative, data analysis cannot allow the accurate establishment of cause-effect relationships, but at least it can support the definition of propositions amongst the constructs of the theory. On the other hand, the explanations of the theory have their origins associated with the observations made by the researcher along the interventions, interpreted by one's theoretical and personal values (i.e., personal, expertise-based, and propositional knowledge).

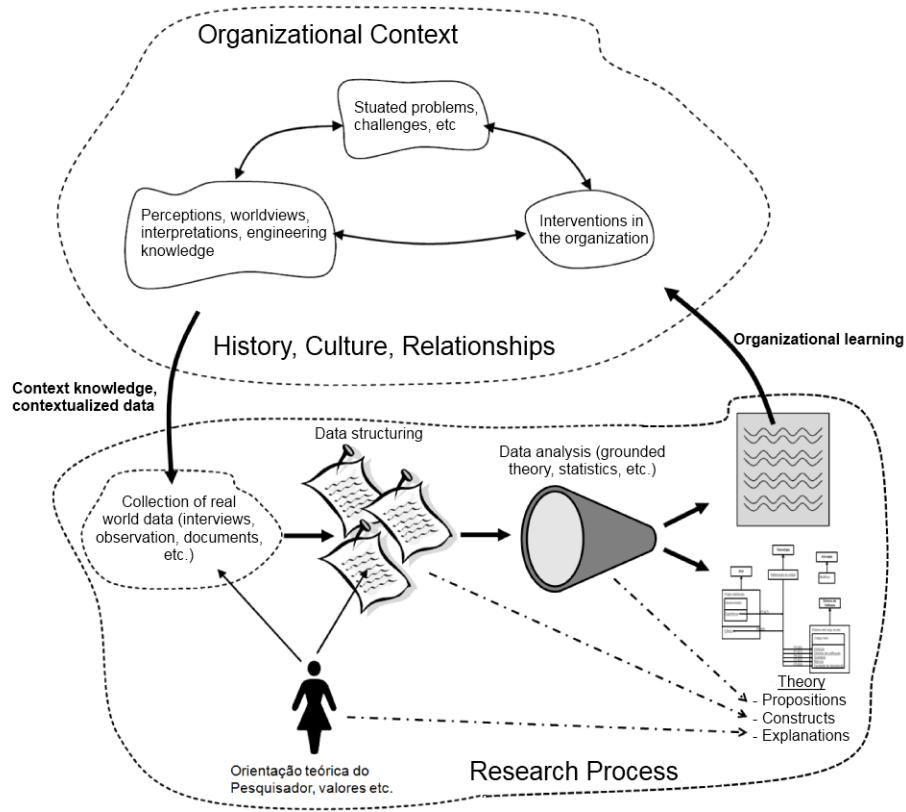

Figure 14 – Action Research process in theory building – evolved from McKay and Marshall (2007)

Due to the way in which theories are built in Action Research, they tend to be of level 1 (minor working relationships that are concrete and directly based on observations) or level 2 (theories of the middle range involving some abstractions but that are still closely linked to observations). The theory built on the refactoring study can be classified on level 1. Theory building in Software Engineering represents a recent research endeavour for several researchers who are trying to offer some sort of guidelines to this end (e.g., Sjøberg *et al.*, 2008). Thus, from what we discussed in this section, it seems that Action Research is suitable for constructing and evaluating Software Engineering theories, but special attention should be given to research rigour, so that relevant results can be produced.

### 5.2.2. Research Rigour and the Results Relevance

The relevance of Action Research results to Software Engineering practice is clear. However, no results are indeed relevant if the means by which they were obtained are not legitimate. To demonstrate result legitimacy it is necessary that a minimum level of rigour is present when conducting the research, which can show that the effects described in the explanations (of the theories) were obtained as an effect of the proposed solution.

Recommendations and strategies to give rigour to Action Research have been widely discussed in the technical literature (Avison *et al.*, 2001, Davison *et al.*, 2004). There are at least two fundamental aspects that can characterize the rigour of a study. One is the control level applied when conducting the study, aimed at minimizing researcher bias and the influence of other variables over the outcomes. The previously mentioned papers address this issue. The second is how theoretical knowledge is kept explicit during the research actions (Baskerville and Pries-Heje, 1999). To address this latter issue, in the absence of approaches in technical literature, we worked on the Action Research planning and analysis stages using the GQM approach (Basili *et al.*, 1994).

The GQM approach is based on the supposition that, in order to effectively measure, goals should be established so that they can guide the preparation of questions to direct the definition of measurements for a particular context. The GQM is set by two processes (Latum *et al.*, 1998): a top-down refinement from the objectives to questions and then in measurements and a bottom-up analysis from the data collected to evaluate them as regards the goals established.

We have used the GQM approach in all Action Research studies that we conducted. We slightly modified the GQM nomenclature to better fit the defined report study format and the Action Research characteristics. So, initially, the research goals are set at the conceptual level, considering the object of study, the point-of-view adopted and the context of the study. Then, the research questions are defined on the practical level where the object of study is characterized according to a quality aspect that is chose for investigation. Lastly, the operational questions determine what data should be collected, keeping the track of the research questions and goals. The 'operational questions' expression is used in place of metrics due to the Action Research capability of using quantitative and qualitative data.

We were able to find in our Action Research studies that, when using GQM, the planning stage becomes more focused because data interpretation is dealt with in advance. Furthermore, it is possible to track the outcomes to the initial goals defined for the study, conferring it an improved rigour. For instance, in the refactoring study GQM was useful in mapping the *grounded theory* categories and research questions, supporting the building of the theory.

Hence, in the same way the GQM approach can link the metrics to the research questions and goals in quantitative studies, it can be used in Action Research studies to keep track of the manipulated tacit knowledge and explicit its use when explaining the observed phenomena. Based on this experience we suggest the combined use of the Action Research and GQM approach as shown in the refactoring study. It will be described in the next section.

### 5.2.3. An Action Research Template Study Report

Despite the vast material found in the technical literature on Action Research we could not find any template to report general Action Research studies, including Software Engineering. Hence, in this section, we propose a template to report Action Research studies (Table 16), derived from our experiences in conducting different Action Research studies in Software Engineering. Although simple, the template is heavily connected with the Action Research process (Section 2.2), presenting one section for each correspondingly process stage. In the sequence each section is going to be explained.

Table 16 - Action Research Report Template

| | | |
|---|---|---|
| **1)** | | **Diagnosis** |
| | a) | Problem Description |
| | b) | Project Context |
| | c) | Research Theme |
| **2)** | | **Planning** |
| | a) | Literature Survey |
| | | i) *Initial Study (Optional)* |
| | b) | **Action Focus** |
| | | i) Objectives |
| | | ii) Research Questions |
| | | iii) Expected Outcomes |
| | c) | **Hypotheses** |
| | d) | **Operational Definitions** |
| | | i) *Techniques* |
| | | ii) *Tools* |
| | | iii) *Instruments* |
| | | iv) *Study Design (Optional)* |
| **3)** | | **Actions** |
| **4)** | | **Evaluation and Analysis** |
| **5)** | | **Reflections and Learning** |
| | a) | Learning |
| | b) | Reflections |

*5.2.3.1. Diagnosis*

The description of the diagnosis stage was split into three sections: problem description, project context, and research theme. Problem description describes the problem faced so that its importance can be remarked. This description should then be contextualized, showing where it is happening, in order to complete diagnosis description. Finally, the research theme section summarizes the problem that will be addressed, linking to the following sections.

*5.2.3.2. Planning*

The planning section is one of the most detailed sections as it supports research execution. Based on the diagnosis previously made the first step intends to describe the technical literature survey that will ground the planning. It should indicate the important aspects of the studies that will be used through the interventions. If the current results of these studies are still being developed or are premature, it is possible to execute an initial (small-scale) controlled study to better evaluate the software technology and thus minimize the risks of applying it in the real scenario. This initial study should also be described in the planning section. Next, the 'action focus' section defines the research goals through the use of the GQM approach. From the establishing of the research goals some hypotheses are determined, showing some expectations on the behaviours of the environment during the interventions. Lastly, the operational definitions should be considered, such as the tools that will be used, analysis techniques, and any other resources necessary to research execution. Optionally, the researcher can also plan how the participants will be organized during the activities, which is common for controlled studies.

*5.2.3.3. Actions*

This section put the interventions in a chronological way, describing how the activities were performed during the research. One basic rule: the more details the better it will be. Special attention should be given to the description of administrative or organizational issues such as the impossibility to execute a planned activity because of an intervention by upper management or the lack of resources.

*5.2.3.4. Evaluation and Analysis*

The goal in this section is to describe the data analysis process and its findings. It is important, when describing the data analysis to try and keep an explicit link between the results obtained from the collected data and the initial objectives. This will preserve the traceability of the outcomes to the diagnosed problem and will assist in giving rigour to the research, as previously discussed.

*5.2.3.5. Reflections and Learning*

The last section has a twofold goal, as Action Research does. First, it is necessary to explore the results achieved against the state-of-the-art found in technical literature. During this examination, the studies used in the 'technical literature survey' section should be emphasized to compare, if possible, the outcomes of the studies.

In a second part, the description of the learning process should not mention only the 'physical' material generated in the context of, and for, the organization, but should also try to depict the learning experience of the participants and how it influenced the organizational culture.

## 6. Final Considerations

In this chapter we described our experience in applying the Action Research methodology in Software Engineering. We extracted a set of recommendations and practices we think can be useful to other researchers interested in using Action Research to investigate real-world software engineering issues where some new software technology is being adopted, or that should be tailored to a specific environment, or even be built from scratch, supported by state-of-the-art corresponding knowledge.

The Action Research methodology is regarded 'as the most realistic research setting found because the setting of the study is the same as the scenario the results will be applied in for a given organization, apart from the presence of the researcher(s)' (Sjøberg *et al.*, 2007). Furthermore we could see that its features are quite suited to dealing with the social component in software engineering practice. All of this is strengthened by the growing interest of the software engineering community in the Action Research methodology as we could find out in a technical literature survey on the use of Action Research in the field. However, it should also be noted that the use of Action Research is not always followed by rigour.

Thus, the Action Research methodology is an appealing alternative to contribute to the domain of Software Engineering research, offering the possibility to conduct studies in new situations. This can lead to an additional number of results with increased relevance as Action Research ensures direct access to the know-how of the practitioners which, in many surveys and controlled studies, for example, is not attainable. As stated by Polanyi (1966): 'we know more than we can say'.

## Acknowledgements

The authors would like to thank the CNPq, CAPES, and FAPERJ agencies for their support to the ESE Group at COPPE research projects. This research is inserted into the context of the eSEE - Experimental Software Engineering Environment. These results could not have been attained without the collaboration of the SIGIC Development Team and the COPPETEC Foundation.

## References

Abrahamsson, P., Koskela, J. (2004) 'Extreme Programming: A Survey of Empirical Data from a Controlled Case Study'. In: Proc. of International Symposium on Empirical Software Engineering, pp. 73-82.

Argyris, C., Putnam, R., Smith, D.M. (1985) 'Action Science'. In: Jossey-Bass Social and Behavioural Science Series, 1st edition.

Avison, D.E., Baskerville, R., Myers, M. (2001) 'Controlling action research projects'. In: Information Technology and People, 14, 28–45.

Avison, D. E., Lau, F., Myers, M. D., and Nielsen, P. A. (1999). 'Action Research'. Communications of the ACM, 42(1):94–97.

Argyris, C., Putnam, R., Smith, D.M. (1985) 'Action Science'. In: Jossey-Bass Social and Behavioural Science Series, 1st edition.

Ballantyne, D. (2004) 'Action Research reviewed: a market-oriented approach'. In: European Journal of Marketing, Vol. 38 No.3/4, pp.321-37.

Basili, V. (1996) 'The role of experimentation: Past, current, and future'. In Proceedings of the 18th International Conference on Software Engineering, pages 442–450.

Basili, V.R., Caldiera, G., Rombach, H.D. (1994) 'The Goal Question Metric Approach'. In the Encyclopaedia of Software Engineering, vol. 2, pp. 528-532, John Wiley & Sons, Inc.

Basili, V., Elbaum, S. (2006) 'Empirically Driven SE Research: State of the Art and Required Maturity'. Invited Talk, ICSE 2006, Shanghai.

Basili, V.R., Selby, R.W., Hutchens, D.H. (1986). 'Experimentation in Software Engineering'. IEEE Transactions on Software Eng., pp. 733-743, July.

Baskerville, R. L. (1999) 'Investigating Information Systems with Action Research'. In: Communications of the Association for Information Systems, volume 2.

Baskerville, R. (2007) 'Educing Theory from Practice'. In N. Kock (Ed.) Information Systems Action Research: An Applied View of Emerging Concepts and Methods. Springer, New York.

Baskerville, R., Pries-Heje, J. (1999). 'Grounded Action Research: A method for understanding IT in practice'. In: Accounting, Management and Information Technologies, 9:1-23.

Baskerville, R., Wood-Harper, A.T. (1996) 'A critical perspective on Action Research as a method for information systems research'. In: Journal of Information Technology, 11, pp. 235–246.

Baskerville, R., Wood-Harper, A. T. (1998) 'Diversity in Information Systems Action Research Methods'. In: European Journal of Information Systems, (7) 2, pp. 90-107.

Bennet, K. H. and Rajlich, V. T. (2000). 'Software maintenance and evolution: A roadmap'. In: The Future of Software Engineering, Ed. ACM Press, 73–87.

Boehm, B.W. (1976). 'Software Engineering,' IEEE Trans. Computers, pp. 1.226 -1.241.

Burns, A. (2005). 'Action research: An evolving paradigm?'. Language Teaching 38(2), pp. 57–74.

Carr, W. (2006). 'Philosophy, Methodology and Action Research'. Journal of Philosophy of Education, Special Issue: Philosophy, Methodology and Educational Research Part 1, 40.2, pp. 421–437.

Carr, W., Kemmis, S. (1986). 'Becoming Critical: Education, Knowledge and Action Research'. Basingstoke: Falmer Press.

Charters, S. M., Budgen, D., Turner, M., Kitchenham, B., Brereton, P., Linkman, S. G. (2009). 'Objectivity in Research: Challenges from the Evidence-Based Paradigm'. Australian Software Engineering Conference, pp. 73-80.

Checkland, P., Holwell, S. (1998). 'Action Research: Its Nature and Validity'. In: Systemic Practice and Action Research 11(1): 9-21.

Crupi, J., Alur, D., Malks, D. (2001). 'Core J2EE Patterns: Best Practices and Design Strategies'. Prentice Hall PTR.

Davison, R. M., Martinsons, M. G., Kock, N. (2004). 'Principles of canonical action research'. In: Information Systems Journal, 14(1), pp. 65–86.

Dick, B. (2004). 'Action Research literature: Themes and trends'. In: Action Research, v. 2, pp. 425-444.

Dunsmore, A., Roper, M., and Wood, M. (2003). 'Practical code inspection techniques for object-oriented systems: an experimental comparison'. In: IEEE Software, 21-29.

Easterbrook, S., Singer, J., Storey, M-A., Damian, D. (2008). 'Selecting empirical methods for software engineering research'. In: Advanced Topics in Empirical Software Engineering, F. Shull, J. Singer, and D.I.K. Sjøberg, eds. Springer-Verlag.

Falbo, R.A., Menezes, C.S., Rocha, A.R.C. (1998). 'A Systematic Approach for Building Ontologies', in Proceedings of the IBERAMIA'98, Lisbon, Portugal.

Fowler, M., Beck, K., Brant, J., Opdyke, W. and D. Roberts. (1999).'Refactoring: Improving the Design of Existing Code'. Addison Wesley, 1st edition.

Greenwood, D. J., Levin, M. (1998). 'Introduction to Action Research: Social Research for Social Change'. Thousand Oaks, CA: Sage Publications.

Guba, E. G., Lincoln, Y. S. (1994). 'Competing paradigms in qualitative research'. In: N. K. Denzin & Y. S. Lincoln (Eds.), Handbook of qualitative research (pp. 105-117). Thousand Oaks, CA: Sage.

Harrison, R., Badoo, N., Barry, E., Biffl, S., Parra, A., Winter, B., Wuest, J. (1999). 'Directions and Methodologies for Empirical Software Engineering Research'. In: Empirical Software Engineering, 4(4), 405–410.

Hathaway, R. S. (1995). 'Assumptions underlying quantitative and qualitative research: implications for institutional research'. In: Research in Higher Education 36(5):535–562.

Healy, M., Perry, C. (2000). 'Comprehensive criteria to judge validity and reliability of qualitative research within the realism paradigm'. Qualitative Market Research: An International Journal, Vol. 3 No.3, pp.118-26.

Heatwole, C. G., Keller, L. F., Wamsley, G. L. (1976). 'Action Research and Public Policy Analysis: Sharpening the Political Perspectives of Public Policy Research'. In: Political Research Quarterly, Vol. 29, pp. 597 – 609.

Hearn, G., Foth, M. (2005). 'Action Research in the Design of New Media and ICT Systems'. In K. Kwansah-Aidoo (Ed.), Topical Issues in Communications and Media Research, New York, NY: Nova Science.

Higgs, J., Burns, A., Jones, M. (2001). 'Integrating clinical reasoning and evidence-based practice'. AACN Clinical Issues 12, pp. 482–490.

Höfer, A., Tichy, W.F. (2007). 'Status of Empirical Research in Software Engineering'. In: Basili et al. (eds), Experimental Software Engineering Issues: Assessment and Future Directions, Springer-Verlag, LNCS 4336.

Holter, I.M., Schwartz-Barcott, D. (1993). 'Action Research: What is it? How has it been used and how can it be used in nursing?'. In: Journal of Advanced Nursing, pp. 298-304.

IEEE Keyword Taxonomy. (2002). http://www2.computer.org/portal/web/publications/acmsoftware.

Juristo, N., Moreno, A. M. (2001). 'Basics of Software Engineering Experimentation'. Kluwer Academic Publisher, USA.

Kampenes, B., Dybå, T., Hannay, J.E., Sjøberg, D.I.K. (2009). 'A systematic review of quasi-experiments in software engineering', Inform. Software Technol. 51 (1), pp. 71-82.

Kates, S., Robertson, J. (2004). 'Adapting Action Research to marketing'. European Journal of Marketing, Vol. 38 No. 3/4, pp.418-32.

Kazman, R., Bass, L. (2002). 'Making Architecture Reviews Work in the Real World,' IEEE Software, vol. 19, No. 1, pp. 62-73.

Kitchenham, B. A., Dybå, T., and Jørgensen, M. (2004). 'Evidence-Based Software Engineering'. In: ICSE 2004, 273-281, IEEE Computer Society Press.

Kitchenham, B. (2007). 'Empirical Paradigm – The Role of Experiments' In: Basili et al. (eds), Experimental Software Engineering Issues: Assessment and Future Directions, Springer-Verlag, LNCS 4336.

Kitchenham, B., Budgen, D., Brereton, P., Turner, M., Charters, S., Linkman, S., (2007). 'Large-scale software engineering questions - expert opinion or empirical evidence?'. In: IET Software, vol.1, no.5, pp.161-171.

Krauss, S.E. (2005). 'Research Paradigms and Meaning Making: A Primer'. In The Qualitative Report, 10 (4), 758-770.

Latum, F., Solingen, R., Hoisl, B., Oivo, M., Rombach, H.D., Ruhe, G. (1998). 'Adopting GQM-based measurement in an industrial environment'. In: IEEE Software, pp. 78-86.

Lau, F. (1997). 'A Review On The Use of Action Research in Information Systems Studies'. In: A. Lee, J. Liebenau, and J. DeGross, (eds.) Information Systems and Qualitative Research, Chapman & Hall, pp. 31-68.

Ludema, J. D., Cooperrider, D. L., Barret, F. J. (2001). 'Appreciative Inquiry: the Power of the Unconditional Positive Question'. In P. Reason & H. Bradbury (Eds.), Handbook of action research: Participative inquiry and practice (pp. 1–14). Thousand Oaks, CA: Sage.

Mäntylä, M.V. (2005). 'An experiment on subjective evolvability evaluation of object-oriented software: explaining factors and interrater agreement', In: International Symposium on Empirical Software Engineering, pp. 277-286.

Mäntylä, M. V. and Lassenius, C. (2006). 'Drivers for software refactoring decisions', In: Proceedings of the 2006 ACM/IEEE International Symposium on Empirical Software Engineering (ISESE'06), pages 297-306, New York, NY, USA.

McKay, J., Marshall, P. (2007). 'Driven By Two Masters, Serving Both'. In N. Kock (Ed.) Information Systems Action Research: An Applied View of Emerging Concepts and Methods. Springer, New York.

Mens, T. and Tourwe, T. (2004). 'A survey of software refactoring'. In: IEEE Transactions on Software Engineering, 30(2):126-139.

Meyer, J. E. (1993). 'New paradigm research in practice: The trials and tribulations of action research'. In: Journal of Advanced Nursing, 18(7), 1066-1072.

Näslund, D. (2002). 'Logistics needs qualitative research – especially Action Research'. In: International Journal of Physical Distribution & Logistics Management, Vol. 32 No. 5, pp. 321-328.

Ottosson, S. (2003). 'Participation action research: a key to improved knowledge of management'. In: The International Journal of Technological Innovation, Entrepreneurship and Technology Management (Technovation), Vol. 23, No. 2, pp. 87-94.

Parnin, C. and Görg, C. (2008). 'A catalogue of lightweight visualizations to support code smell inspection', In: Proceedings of the ACM Symposium on Software Visualization, pp. 77-86.

Pfleeger, S. H. (1999). 'Albert Einstein and empirical software engineering'. IEEE Computer, 32(10), 32–37.

Polanyi, M. (1966). 'The Tacit Dimension'. Routledge and Keoan Paul, London.

Rainer, A., Jagielska, D. and Hall, T. (2005). 'Software Practice versus evidence-based software engineering research'. In: Proceedings of the Workshop on Realising Evidence-based Software Engineering, ICSE-2005.

Ramesh, V., Glass, R.L., and Vessey, I. (2004). 'Research in computer science: an empirical study'. Journal of Sys. And Sw., 165-176.

Reason, P., Bradbury, H. (2001). 'Introduction: Inquiry and participation in search of a world worthy of human aspiration'. In P. Reason & H. Bradbury (Eds.), Handbook of action research: Participative inquiry and practice (pp. 1-14). Thousand Oaks, CA: Sage.

Rus, I., Lindvall, M. (2002). 'Knowledge Management in Software Engineering' IEEE Software (19:3), pp. 26-38.

Ruthruff, J. R., Penix, J., Morgenthaler, J. D., Elbaum, S., Rothermel, G. (2008). 'Predicting accurate and actionable static analysis warnings: an experimental approach', In: Proc. of the 30th International Conference on Software Engineering, pp. 341-350.

Santos, P.S.M., Travassos, G. H. (2009). 'Action Research Use in Software Engineering: an Initial Survey' 3nd International Symposium on Empirical Software Engineering and Measurement, pp. 414-417, Orlando, USA.

Santos, P.S.M., Travassos, G.H. (2010). 'Quality Inspection in Use Case Descriptions: an Experimental Evaluations in a Real Project', IX Brazilian Symposium of Software Quality, pp. 261-275, Belém, Brazil. (in Portuguese)

Seaman, B.C. (1999). 'Qualitative Methods in Empirical Studies of Software Engineering'. In: IEEE Transactions on Software Engineering, vol. 25(4), pp. 557-572.

Schein, E.H. (1994). 'The process of dialogue: creating effective communication'. In: The Systems Thinker, Vol. 5 No. 5, pp. 1-4.

Shull, F., Carver, J., Travassos, G.H.(2001). 'An Empirical Methodology for Introducing Software Processes'. In: 8th European Software Engineering Symposium and 9$^{th}$ ACM SIGSOFT Symposium on the Foundations of Software Engineering (FSE-9) and 8$^{th}$ European Software Engineering Conference (ESEC), Vienna, Austria, September.

Sjøberg, D. I. K., Dybå, T., Anda, B.C.D. and Hannay, J. E. (2008). 'Building Theories in Software Engineering'. Advanced Topics in Empirical Software Engineering, F. Shull, J. Singer, and D.I.K. Sjøberg, eds. Springer-Verlag.

Sjøberg, D.I.K., Dybå, T., Jørgensen, M. (2007). The future of empirical methods in software engineering research. In FOSE '07: Future of Software Engineering, pages 358-378, Washington, DC, USA,.

Sjøberg, D.I.K., Hannay, J.E., Hansen, O., Kampenes, V.B., Karahasanović, A., Liborg, N.-K., Rekdal, A.C. (2005). 'A Survey of Controlled Experiments in Software Engineering'. IEEE Transactions on Software Engineering, 31(9): 733-753.

SOFTEX (2007). 'MPS.BR: Brazilian Software Process Improvement', General Guide Version 1.2, Campinas, SP, SOFTEX.

Strauss, A., Corbin, J. (1990). Basics of qualitative research: Grounded theory procedures and techniques. Newbury Park, CA: Sage.

Sun (1999). 'Code Conventions for the Java Programming Language', Sun Microsystems. Available at: http://java.sun.com/docs/codeconv/. Accessed in: July 2010.

Susman, G.L., Evered, R.D. (1978). 'An assessment of the scientific merits of Action Research'. In: Administrative Sciences Quarterly, 23, pp. 582–603.

Thiollent, M. (2007). 'Action Research Methodology', Cortês Editora, 15$^{th}$ Ed. (in Portuguese).

Tichy, W. F. (1998). 'Should Computer Scientists Experiment More?' IEEE Computer, pp. 32-40, May.

TRAC (2003). 'TRAC: Integrated Software Configuration and Project Management' Available at: http://trac.edgewall.org/. Accessed in: July 2010.

Travassos, G. H. and Kalinowski, M. (2009). 'iMPS 2009 : characterization and performance variation of software organizations that adopted the MPS model'. Association for Promotion of the Excellence the Brazilian Software - SOFTEX, 2009 . ISBN 978-85-99334-18-8. Available at http://www.softex.br/mpsbr/_livros/arquivos/Softex%20iMPS%202009%20Ingles_vFinal_12jan10.pdf. Accessed in: August 2010

Tripp, D. (2005). 'Action Research: A methodological introduction,' Educação e Pesquisa, (31:3), pp. 443-466.

Westbrooke, R. (1995). 'Action Research: a new paradigm for research in production and operations management', International Journal of Operations and Production Management, Vol. 15, No.12, pp 6-20.

Wöhlin, C., Runeson, P., Höst, M., Ohlsson, M. C., Regnell, B., and Wesslén, A. (2000). 'Experimentation in Software Engineering: An Introduction'. Kluwer Academic Publishers.

Wöhlin, C., Runeson, P., Höst, M., Ohlsson, M. C., Regnell, B., and Wesslén, A. (2000). 'Experimentation in Software Engineering: An Introduction'. Kluwer Academic Publishers.

Wöhlin, C., Höst, M., Henningsson, K. (2003). 'Empirical Research Methods in Software Engineering'. In Lecture Notes in Computer Science: Empirical Methods and Studies in Software Engineering: Experiences from ESERNET, edited by A. I. Wang and R. Conradi, Springer Verlag.

Zelkowitz, M.V. (2007). 'Techniques for Empirical validation' In: Basili et al. (eds), Experimental Software Engineering Issues: Assessment and Future Directions, Springer-Verlag, LNCS 4336.

## Technical Literature Survey References

Abrahamsson, P., Koskela, J. (2004). 'Extreme Programming: A Survey of Empirical Data from a Controlled Case Study'. In: Proc. of International Symposium on Empirical Software Engineering, pp. 73-82.

Bengtsson, P., Bosch, J. (1999). 'Haemo Dialysis Software Architecture Design Experiences'. Proc. 21st ICSE, pp. 516-525.

Bosch, J. (2010). 'Toward Compositional Software Product Lines'. IEEE Software, vol.27, no.3, pp. 29-34, May-June.

Canfora, G., Garcia, F., Piattini, M., Ruiz, F., Visaggio, C.A. (2006). 'Applying a framework for the improvement of the software process maturity in a software company'. Journal Software Practice and Experience 36 (3) 283-304.

Fernández-Medina, E. and Piattini M. (2005). 'Designing secure databases'. In: Information & Software Technology 47(7), pp. 463-477.

Fitzgerald, B., O'Kane, T. (1999). 'A Longitudinal Study of Software Process Improvement'. IEEE Software, 16, 3, pp. 37-45.

Gutierrez, C., Rosado, D. G., Fernandez-Medina, E. (2009). 'The practical application of a process for eliciting and designing security in web service systems'. Information and Software Technology, Volume 51, Issue 12, pp. 1712-1738.

Kauppinen, M., Vartiainen, M., Kontio, J., Kujala, S., Sulonen, R. (2004). 'Implementing requirements engineering processes throughout organizations: success factors and challenges'. In: Information and Software Technology, vol. 46, pp. 937-953.

Kautz, K, Hansen, H.W., Thaysen, K. (2000). 'Applying and Adjusting a Software Process Improvement Model in Practice: The Use of the IDEAL Model in a Small Software Enterprise'. Proc. 22nd Int'l Conf. Software Eng., IEEE CS Press, pp. 626-633.

Lindvall, M. and Sandahl, K. (1996). 'Practical Implications of Traceability'. Journal of SP&E, 26(10):1161-1180, October.

Lycett, M. (2001). 'Understanding 'Variation' in component-based development: Case findings from practice'. Information and Software Technology, 43(3), 203-213.

Maiden, N., Robertson, S. (2005). 'Developing use cases and scenarios in the requirements process'. Proc. ICSE 2005, 561-570.

Mattsson, A., Lundell, B., Lings, B., Fitzgerald, B. (2009). 'Linking Model-Driven Development and Software Architecture: A Case Study Software Engineering'. IEEE Transactions on 35, 83-93.

McCaffery, F., Burton, J., Richardson, I. (2009). 'Improving software Risk Management in a Medical Device Company' 31st International Conference on Software Engineering, pp.152-162.

Napier, N.P., Mathiassen, L., Johnson, R.D. (2009). 'Combining Perceptions and Prescriptions in Requirements Engineering Process Assessment: An Industrial Case Study'. IEEE Transactions on Software Engineering, vol.35, no.5, pp.593-606.

Nielsen, P. A., Tjørnehøj, G. (2010). 'Social networks in software process improvement'. Journal of Software Maintenance and Evolution: Research and Practice, vol. 22, no. 1, pp. 33-51.

Pino, F. J., Pardo, C., Garcia, F., Piattini, M. (2010). 'Assessment methodology for software process improvement in small organizations'. Information and Software Technology, In Press, Corrected Proof, Available online 1 May 2010, ISSN 0950-5849.

Polo, M., Piattini, M., Ruiz, F., (2002). 'Using a Qualitative Research Method for Building a Software Maintenance Methodology'. In: Software Practice & Experience, 32(13), pp. 1239-1260.

Salo, O., Abrahamsson, P. (2005). 'Integrating Agile Software Development and Software Process Improvement: a Longitudinal Case Study'. ISESE, Australia, Noosa Heads.

Staron, M., Meding, W. (2008). 'Predicting weekly defect inflow in large software projects based on project planning and test status'. Information and Software Technology.

Staron, M., Meding, W., Nilsson, C. (2009). 'A framework for developing measurement systems and its industrial evaluation'. In: Information and Software Technology, 51(4):721-737, April.

Vigder, M., Vinson, N. G., Singer, J., Stewart, D., Mews, K. (2008). 'Supporting Scientists' Everyday Work: Automating Scientific Workflows'. IEEE Software, 25, 52-58.

# ON THE AUTHORS

**Paulo Sérgio Medeiros dos Santos** is a doctorate student at COPPE – Federal University of Rio de Janeiro – Brazil where he received his Master's degree in 2009. His current research interests include experimental software engineering, research methodologies and theory building, cognitive science, and artificial intelligence, applied to software engineering. He has industrial experience in software development as a developer and architect, and currently works at a major Brazilian oil company. Most of the information regarding his research projects and working activities can be found at http://ese.cos.ufrj.br/ese.  He can be contacted at pasemes@cos.ufrj.br.

**Guilherme Horta Travassos** is a Professor of Software Engineering with the Systems Engineering and Computer Science Department at COPPE/Federal University of Rio de Janeiro - Brazil. He is also a 1D CNPq - Brazilian Research Council researcher. He received his doctorate degree from COPPE/UFRJ in 1994 and spent 2 years with the Experimental Software Engineering Group at the University of Maryland - College Park, in a post-doctoral position (1998/2000). He leads the Experimental Software Engineering Group at COPPE/UFRJ. His current research interests include experimental software engineering, e-science and non-conventional Web applications, software quality, and VV&T concerned with object-oriented software. He is a member of ISERN, ACM and SBC (Brazilian Computer Society).   He can be contacted at ght@cos.ufrj.br.